\documentclass{aas}
\usepackage{graphicx}
\usepackage{tabularx}
\usepackage{longtable}
\usepackage{rotating}

\received{  }
\revised{}
\accepted{}
\submitjournal{ } 

\shorttitle{Search for hot subdwarf stars in Gaia DR2 }
\shortauthors{Lei et al.}

\begin{document}

\title{Hot subdwarf stars identified in Gaia DR2 with spectra of LAMOST DR6 and DR7 I. Single-lined spectra}

\correspondingauthor{Gang Zhao}
\email{gzhao@nao.cas.cn}

\correspondingauthor{Zhenxin Lei}
\email{zxlei@nao.cas.cn}

\author{Zhenxin Lei}
\affiliation{Physics Department, Xiangtan University, Xiangtan 411105, China\\}
\affiliation{Key Laboratory of Optical Astronomy, National Astronomical Observatories, 
Chinese Academy of Sciences, Beijing 100012, China\\}

\author{Jingkun Zhao }
\affiliation{Key Laboratory of Optical Astronomy, National Astronomical Observatories, 
Chinese Academy of Sciences, Beijing 100012, China\\}

\author{P\'eter N\'emeth}
\affiliation{ Astronomical Institute of the Czech Academy of Sciences, CZ-251\,65, Ond\v{r}ejov, Czech Republic\\}
\affiliation{ Astroserver.org, 8533 Malomsok, Hungary\\
}

\author{Gang Zhao }
\affiliation{Key Laboratory of Optical Astronomy, National Astronomical Observatories, 
Chinese Academy of Sciences, Beijing 100012, China\\}



\begin{abstract}
182 single-lined hot subdwarf stars are identified by using spectra from the 
sixth and seventh data release (DR6 and DR7) of the Large Sky Area Multi-Object 
Fibre Spectroscopic Telescope (LAMOST) survey. We classified all the 
hot subdwarf stars using a canonical classification scheme, and got 
89 sdB, 37 sdOB, 26 sdO, 24 He-sdOB, 3 He-sdO and 3 He-sdB stars, respectively. 
Among these stars, 108 hot subdwarfs are newly discovered, while 74 stars 
were reported by previous catalogs. The atmospheric parameters of these stars 
were obtained by fitting the hydrogen (H) and helium (He) lines with non-local 
thermodynamic equilibrium (non-LTE) model atmospheres.  
The atmospheric parameters confirm the two He sequences and the two subgroups of He-sdOB stars 
in our samples, which were found by previous studies in the 
$T_{\rm eff}$-$\mathrm{log}(n\mathrm{He}/n\mathrm{H})$ diagram. 
Our results demonstrate different origins of field hot subdwarf stars and 
extreme horizontal branch (EHB) stars in globular clusters (GCs), and provide 
strict observational limits on the formation and evolution models of the different 
sub-types of these evolved objects. Based on the results, 
we evaluated the completeness of the Geier et al. (2019) catalog. We found the 
fraction of hot subwarf stars is between 10\% and 60\%, 
depending on the brightness of the sample. 
A more accurate estimation for the hot subdwarf fraction can be obtained when similar  
results from composite spectra will become available. 

\end{abstract}



\section{Introduction} \label{sec:intro}
Hot subdwarf stars (i.e., sdO/B) are evolved low mass stars around at 0.5 M$_{\odot}$. They show  
similar spectra to main-sequence (MS) stars of O/B type, but at  
lower luminosity and with broader spectral features. 
These hot stars occupy the extreme blue region 
of horizontal branch (HB) in the Hertzsprung-Russell (HR) diagram and 
burn He in their cores, therefore, 
they are also known as extreme horizontal branch (EHB) stars (Heber 1986). 

The formation and evolution of hot subdwarf stars are still 
not clear. To end up on the EHB, the 
progenitors of these hot stars have to lose nearly the whole 
envelope mass by the end of their red giant branch (RGB) stages. Therefore, 
binary evolution is thought to be the main formation mechanism for 
hot subdwarf stars (Han et al. 2002, 2003). On the observational 
side, about half of the hot subdwarfs are 
found in close binaries (Maxted et al. 2001; Napiwotzki et al. 2004; Copperwheat et al. 2011), 
and their companions could be brown dwarfs, MS stars, white dwarfs (WDs), 
and even neutron stars  or black holes (Kupfer et al. 2015; Kawka et al. 2015). 
Studies on hot subdwarf stars therefore can shed light on the details of  
binary evolution processes, such as Roche lobe overflow (RLOF), common
envelope (CE) ejection and the merger of two He WDs  
(Han et al. 2002, 2003; Chen et al. 2013; Zhang \& Jeffery 2012; Zhang et al. 2017; Vos et al. 2019). 
Moreover, close hot subdwarf binaries with 
compact companions (e.g., WDs, neutron stars or black holes) 
are potential verification sources (Kupfer et al. 2018) for the future space based  
gravitational wave (GW) detectors, such as 
LISA (Amaro-Seoane et al. 2017) and TianQin (Luo et al. 2016a). 
Close hot subdwarf + massive WD binaries are possible progenitor systems for type Ia supernovae, in which 
the hot subdwarf companion may survive the explosion as a hypervelocity remnant 
(Wang et al.2009; Vennes et al. 2017; Li et al. 2018; Raddi et al. 2019).   

Pulsating sdO/B stars allow the accurate determination 
of mass and internal structure by using asteroseismic 
methods, which provide excellent tests for the formation and evolution models
(Kawaler et al. 2010; Charpinet et al. 2011; 
Baran et al. 2012; Battich et al. 2018; Zong et al. 2018). 
The diversity of atmospheres in hot subdwarf stars  
makes them good samples to study  atomic diffusion processes (Moehler et al. 2014; Jeffery et al.
2017; N\'emeth 2017; Byrne et al. 2018; Naslim et al. 2013, 2019). In addition, 
hot subdwarf stars in Globular Clusters (GCs) provide useful information  
to understand the formation and evolution of the oldest populations in our 
galaxy (Lei et al. 2015, 2016; Latour et al. 2014, 2018). For recent  
reviews on these special stars see Heber (2009, 2016). 

Since Kilkenny et al. (1988) published the first catalog 
of 1225 spectroscopically identified hot subdwarf stars, the number 
of these special blue objects exploded with the data release of 
many spectroscopic surveys, such as  the Hamburg Quasar Survey (HS,
Hagen et al. 1995), the Hamburg ESO survey (HE, Wisotzki
et al. 1996), the Edinburgh-Cape Survey (EC, Kilkenny et al.
1997),  the Byurakan surveys (FBS, SBS, Mickaelian et
al. 2007, 2008), the GALEX all sky survey (Vennes et al. 2011; N\'emeth et al. 2012), 
the Sloan Digital Sky Survey (SDSS, Geier et al. 2015; Kepler al. 2015, 2016)  
and the LAMOST survey (Luo et al. 2016b, 2019; Lei et al. 2018, 2019a, b, Bu et al. 2017, 2019).  

By retrieving known hot subdwarfs and candidates 
from the literature and unpublished databases, Geier et al. (2017) 
compiled a hot subdwarf catalog with 5613 objects. 
This catalog provide many useful stellar information, such as 
multi-band photometry, proper motions, classifications, atmospheric parameters, etc. 
The second data release of the Gaia mission (Gaia DR2, Gaia Collaboration et al. 2018) brought us 
excellent opportunities to discover new hot subdwarf stars, because it 
provides accurate positions, photometry, parallax and proper motions. 
With these information, one can easily compile a large sample of hot subdwarf candidates with 
high confidence in the HR-diagram. Using this method, 
Geier et al. (2019) compiled a list of 39800 hot subdwarf 
candidates selected from Gaia DR2, which is the largest collection ever published. 
This catalog can be used as a good 
input target list for follow-up spectroscopic analyses. 

In this paper, 
we analyzed the single-lined spectra which were selected by 
cross-matching the catalog of Geier et al. (2019) with the 
latest data release of the LAMOST survey, i.e., DR6 and DR7.  
By employing non-Local
Thermodynamic Equilibrium (non-LTE) model atmospheres, we obtained 
their atmospheric parameters (e.g., effective temperatures, 
surface gravity values and He abundance). 
We also found many composite spectra in the sample, and will 
report the analysis and results in a forthcoming paper. 
The structure of this paper is the following:   
In section 2, we introduced the hot subdwarf candidates catalog and  
the databases of LAMOST DR6 and DR7, and how the candidate spectra were selected. 
The method of spectral analysis and classification for 
hot subdwarf stars are described in section 3. We give our results in 
section 4. We finish the paper with a brief discussion and summary in section 5.

\section{Target selection}
\subsection{The hot subdwarf candidate catalogue} 
\begin{figure}
\centering
\includegraphics [width=110mm]{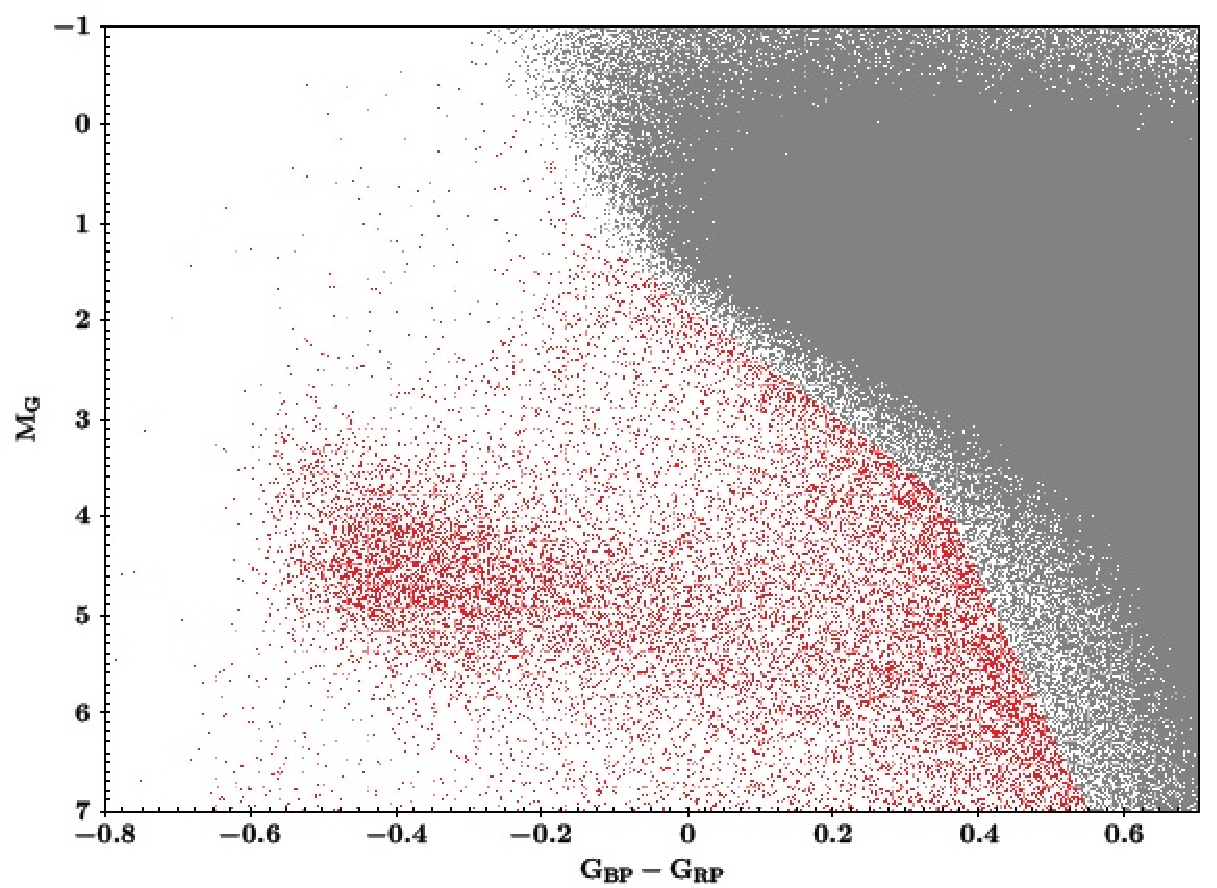}
\caption{Hot subdwarf candidates (red dots) in the Gaia DR2 HR-diagram. 
The figure is taken from Geier et al. (2019).}
\end{figure} 
 
Geier et al. (2019) compiled a catalog of hot subdwarf candidates from Gaia DR2, 
in which 39800 candidates were selected by means of color, absolute magnitude 
and reduced proper motion cuts. Fig 1 shows the selected candidates (red dots) by Geier et al. (2019) 
in the Gaia DR2 HR-diagram (the figure is taken from Fig 3 of Geier et al. 2019, for the 
detailed candidate selection filter please see 
section 3 in their study). 
The majority of the candidates are expected to be 
hot subdwarf stars of spectral type B and O, followed by blue horizontal branch (BHB) stars, 
hot post-AGB stars, and central stars of planetary nebulae.  
 
\subsection{The LAMOST DR6 and DR7 database}  
LAMOST is a Chinese national scientific research facility operated 
by the National Astronomical Observatories,  Chinese Academy of Sciences. 
It has a specially designed Schmidt telescope with 4000 fibers in a field of view of 20 deg$^{2}$ in the sky 
(Cui et al. 2012; Zhao et al. 2006, 2012). LAMOST finished its 
pilot survey in June 2012 and the first-six-years regular survey in July 2018, respectively. The data 
from both surveys make up the sixth data release (DR6) of LAMOST, in which 
9919106  spectra have been obtained in the optical band (e.g., 3690-9100 $\mathrm{\AA}$) with a resolution of 1800 
at 5500 $\mathrm{\AA}$. LAMOST DR6 contains 8966416 stellar spectra, 
172866 galaxy spectra, 60173 quasar spectra, and 719651 spectra of unknown objects, respectively. 
The data obtained in the pilot survey and the first-five-years regular survey (ended in July 2017) make up the 
LAMOST DR5 database. The LAMOST DR6 database contains the whole LAMOST DR5 database and  
new observational data from the sixth year survey (i.e., 889947 spectra,  observed from September 2017 to July 2018). 

The LAMOST seventh year survey also have been completed 
(e.g., from September 2018 to July 2019). The 558412 low resolution 
(e.g., $\lambda/\delta \lambda =$ 1800) spectra observed during this period are released as LAMOST DR7\_v0 database. 
Though the final version of the LAMOST DR7 database,    
which consists of the whole DR6 database and the new spectra observed 
during the seventh year survey,  will be publicly available on March 2020, all the new 
spectra observed during this period are already included in the LAMOST DR7\_v0 database.  
 
\subsection{Hot subdwarf candidates selected by cross-matching the Geier et al.(2019) catalogue with LAMOST DR6 and DR7} 
In our previous work, we have already identified 682 hot subdwarf stars by combining the Gaia DR2 database 
and the LAMOST DR5 database (Lei et al. 2018, 2019b), among which 241 stars were newly discovered. 
In this study, we analyzed hot subdwarf candidates selected from the LAMOST DR6 and DR7\_v0 database. 
We selected the candidates by the following steps:   
First, we cross-matched the Geier et al (2019) catalogue  
with the LAMOST DR6 and DR7\_v0 database separately, and  obtained 
2513 common objects in total. Then, we downloaded all the spectra of the common stars 
from the LAMOST website (www.lamost.org), and 
selected 1348 spectra with signal to noise ratio larger than 10 in the \textit{u} band (SNR-u), 
which guaranties a sufficient quality for spectral analysis. 
After removing the spectra that have been analyzed in 
our previous studies (Lei et al. 2018, 2019a, b), composite spectra\footnote{We will  
analyze composite spectra and report those results in a forthcoming paper.}, 
and  duplicate sources, we finally got 607 spectra that are suitable for 
spectral analysis.

\section{Spectroscopy and spectral classification} 
As did in Lei et al. (2018, 2019b), we employed the spectral analysis tool, {\sc XTgrid} (N\'emeth et al. 2012, 2014) 
to analyze the selected 607 spectra. {\sc XTgrid} fits the observed data with synthetic spectra 
({\sc Synspec} version 49; Lanz et al 2007) 
calculated from non-LTE model atmospheres ({\sc Tlusty} version 204; Hubeny \& Lanz 2017). 
The best fitting model is searched for iteratively 
with a successive approximation method along the steepest-gradient of the $\chi^2$ field. 
Parameter uncertainties have been estimated by mapping the $\Delta\chi^2$ field until the 60 percent 
confidence level at the given number of free parameters was reached. For the detailed information to 
obtain the parameter error bars, the readers are suggested to see Fig 2 in Lei et al. (2019b) and the text therein. 

\begin{figure}
\centering
\begin{minipage}[c]{0.48\textwidth}
\includegraphics [width=85mm]{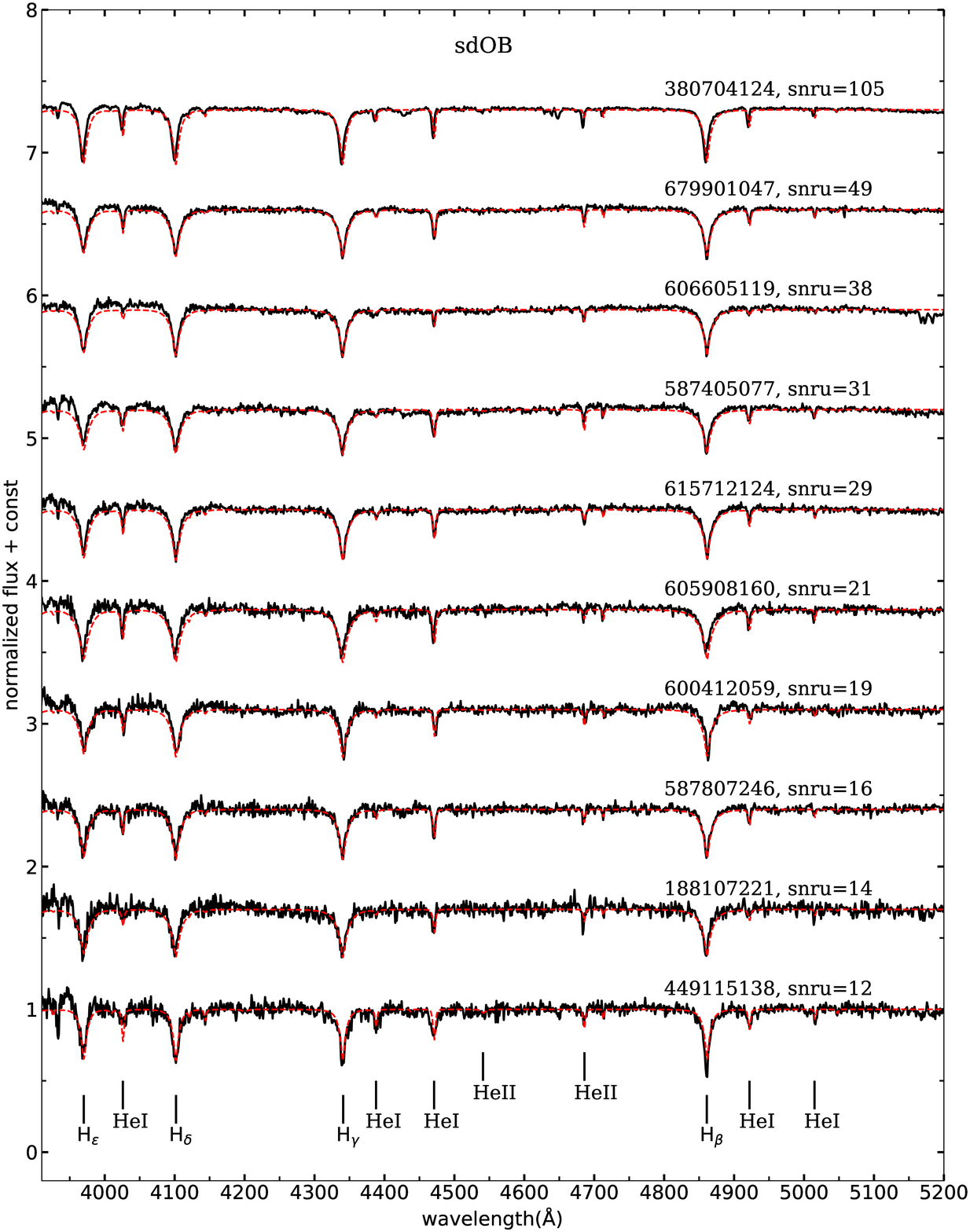}
\end{minipage}%
\begin{minipage}[c]{0.48\textwidth}
\includegraphics [width=85mm]{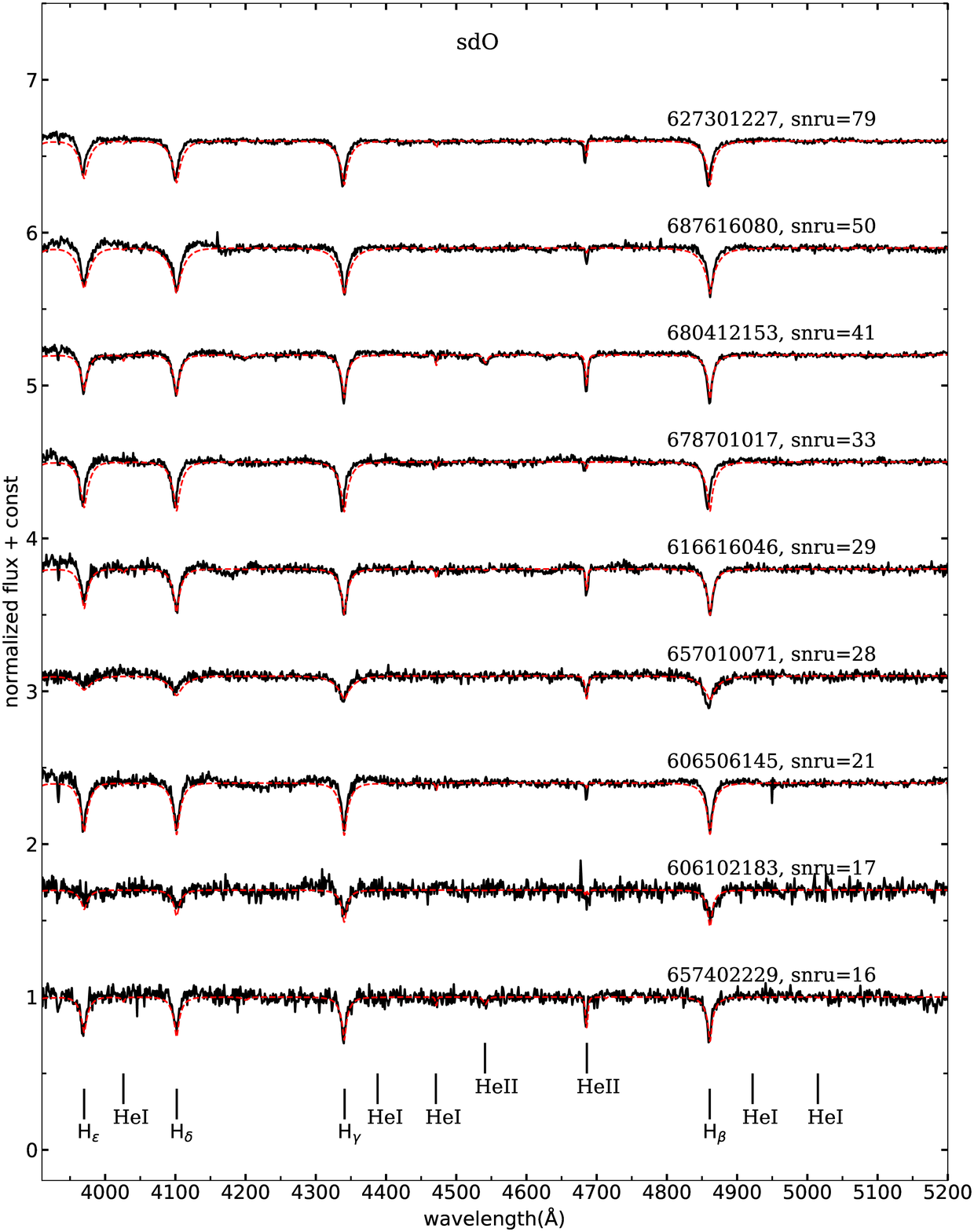} 
\end{minipage}%
\caption{The best-fit models reached by {\sc XTgrid} for some sdOB 
(left) and sdO stars (right). 
The red dashed curves are the best-fitting synthetic spectra, 
while the black solid curves are the observed 
spectra. Some important H and He  
lines in the wavelength range of 3900-5200 $\mathrm{\AA}$ 
are marked by short vertical lines at the bottom of the panels. 
The long integers at the right of the panels 
are the LAMOST obsid for each observed spectrum. 
The SNR in the \textit{u} band for observed spectra increase from bottom to top.}   
\end{figure}

The atmospheric parameters of all candidates, such as effective temperature ($T_{\rm eff}$), 
surface gravity ($\log{g}$) and He abundance ($\log(n{\rm He}/n{\rm H})$), 
are obtained by the method described above. 
Some of the best-fit models for sdOB and sdO stars are shown in Fig 2.  
As did in our previous studies (Lei et al. 2018, 2019b), we identified stars 
with $T_{\rm eff}$ hotter than 20000 K and $\log{g}$ larger than 5.0 $\mathrm{cm\ s^{-2}}$ as 
hot subdwarf stars. On the other hand, the stars with $T_{\rm eff}$ lower than 20000 K or  
$\log{g}$ lower than 5.0 $\mathrm{cm\ s^{-2}}$ are considered as BHB stars or B type Main-sequence (B-MS) stars, while 
for a few stars with very high  $T_{\rm eff}$ and $\log{g}$ 
(e.g., $T_{\rm eff}>$ 80000 K and $\log{g} >$ 7 $\mathrm{cm\ s^{-2}}$), we classified them as 
white dwarf stars (WDs).  We focus only on hot subdwarf stars in the rest of this paper. 

We used the spectral classification scheme of Moehler et al. (1990) and Geier et al. (2017) to 
classify the identified hot subdwarf stars in this study. Stars with dominant H Balmer lines, 
but weak or absent He lines, are classified as sdB stars. 
Stars with dominant H Balmer lines and an obvious 
He II 4686 $\mathrm{\AA}$ line,  but without obvious He I lines are considered as sdO stars. 
Stars having dominant H Balmer lines, and both weak He I and He II lines are identified as 
sdOB stars. Stars with dominant He I lines, but weak or absent H Balmers and He II lines, 
we classified  as He-sdB stars. Stars presenting strong He II lines, but with weak or absent 
H Balmer lines  and He I lines are He-sdO stars, while the stars presenting both strong 
He I and He II line, but with weak or absent H Balmer lines are classified as He-sdOB stars.

\section{Results } 
From the 607 selected candidates, we identified 182 single-lined hot subdwarf stars, 
including 89 sdB stars, 37 sdOB stars, 26 sdO stars, 24 He-sdOB stars, 3 He-sdO stars, and 
3 He-sdB stars. By cross-matching with the hot subdwarf stars cataloged by  
Geier et al. (2017), we found 74 common objects. That means we have found 
108 new, previously uncatalogued hot subdwarf stars and obtained their atmospheric parameters by detailed 
spectral analysis for the first time. 

Table 1 gives the parameters and information for the 182 hot subdwarf stars with 
single-lined spectra. Columns 1-4 give the 
 right ascension (RA), declination (DEC), LAMOST\_obsid,  and Gaia source\_id. 
Columns 5-7 give the atmospheric parameters fitted by {\sc XTgrid}, 
e.g., $T_{\rm eff}$, $\log{g}$ and $\log(n{\rm He}/n{\rm H})$, while 
columns 8-10 give the SNR in the \textit{u} band, the apparent Gaia $G$ band magnitudes and 
spectral classification. The 74 stars common with the hot subdwarf catalog of 
Geier et al. (2017) are marked by $^{*}$.

\begin{table*}
\tiny
 \begin{minipage}{160mm}
  \caption{Information on the 182 hot subdwarf stars identified in this study. From left to right, we list  
  the right ascension (RA), declination (DEC), LAMOST\_obsid,  and Gaia source\_id. Then the $T_{\rm eff}$, $\log{g}$ and $\log(n{\rm He}/n{\rm H})$ are listed from the {\sc XTgrid} fits. Next, the SNR in the $u$ band,  the apparent magnitudes in the Gaia $G$ band and the spectral classifications are listed, respectively.} 
  \end{minipage}\\
  \centering
    \begin{tabularx}{16.0cm}{lllccccccccccccccX}
\hline\noalign{\smallskip}
RA\tablenotemark{a}  & DEC   &obsid & source\_id 
& $T_\mathrm{eff}$ &  $\mathrm{log}\ g$ & $\mathrm{log}(n\mathrm{He}/n\mathrm{H})$\tablenotemark{b}
&SNRU & $G$ & spclass  \\
 LAMOST &  LAMOST& LAMOST&Gaia  &(K)&($\mathrm{cm\ s^{-2}}$)& &&Gaia(mag) & \\
\hline\noalign{\smallskip}
1.8907183$^{*}$ & 13.5993244 & 619614193 & 2767874292175410560 & 29560$\pm$120 & 5.41$\pm$0.01 & -1.90$\pm$0.04 & 103 & 13.07 & sdB \\
2.1021972 & 49.083822 & 593009050 & 393589879591384576 & 26640$\pm$700 & 5.53$\pm$0.08 & -2.61$\pm$0.11 & 17 & 15.91 & sdB \\
2.7184872 & 26.5002178 & 689110148 & 2850670743266825600 & 28380$\pm$190 & 5.27$\pm$0.02 & -2.34$\pm$0.09 & 13 & 16.98 & sdB \\
2.9364712 & 46.801838 & 593007055 & 392942881419391872 & 45110$\pm$420 & 5.34$\pm$0.04 & 0.91$\pm$0.23 & 56 & 14.37 & He-sdOB \\
4.2165006 & 52.146517 & 615603055 & 395157267782903808 & 29170$\pm$350 & 5.46$\pm$0.06 & -2.56$\pm$0.15 & 11 & 16.81 & sdB \\
4.23055 & 51.230486 & 615605186 & 394991241522199040 & 32770$\pm$460 & 5.48$\pm$0.07 & -3.04$\pm$0.58 & 14 & 16.36 & sdB \\
4.6252262 & 48.805384 & 593013047 & 392840562417338112 & 24740$\pm$140 & 5.02$\pm$0.03 & -1.57$\pm$0.06 & 11 & 15.30 & sdB \\
4.7865313 & 52.511876 & 615603207 & 419143904215897728 & 51870$\pm$5990 & 5.19$\pm$0.30 & -1.46$\pm$0.14 & 10 & 17.16 & sdO \\
5.8632988 & 51.130463 & 615605166 & 394843322846749824 & 26610$\pm$170 & 5.42$\pm$0.03 & -2.43$\pm$0.06 & 21 & 15.95 & sdB \\
6.0352547 & 56.027472 & 605908160 & 421328839978415616 & 35020$\pm$560 & 5.76$\pm$0.13 & -1.34$\pm$0.10 & 21 & 16.18 & sdOB \\
6.1375712 & 26.8194824 & 689109217 & 2856144494402348544 & 47370$\pm$710 & 5.82$\pm$0.11 & 0.30$\pm$0.09 & 30 & 16.90 & He-sdOB \\
6.51063$^{*}$ & 31.1057 & 679407014 & 2862194144817359872 & 30150$\pm$110 & 5.52$\pm$0.08 & -2.98> & 83 & 14.87 & sdB \\
7.329288 & 52.97546 & 615609066 & 416403783797286784 & 36080$\pm$490 & 5.66$\pm$0.58 & -2.82> & 12 & 16.80 & sdB \\
15.599966 & 48.879263 & 353516071 & 402544091832710272 & 57260$\pm$9490 & 6.22$\pm$0.78 & 0.66$\pm$1.04 & 28 & 16.67 & He-sdOB \\
15.866097 & 32.675987 & 96304147 & 314344331362996736 & 37490$\pm$1030 & 5.43$\pm$0.08 & -1.69$\pm$0.14 & 14 & 14.37 & sdOB \\
16.358834 & 49.928952 & 686402134 & 404204083809859584 & 53250$\pm$6940 & 5.75$\pm$0.06 & -2.41$\pm$0.18 & 23 & 16.88 & sdO \\
17.418584 & 52.819013 & 686415207 & 404958378847936000 & 35330$\pm$250 & 5.90$\pm$0.08 & -1.47$\pm$0.06 & 14 & 17.52 & sdOB \\
18.320503 & 47.191618 & 603604248 & 401413450281523584 & 48680$\pm$820 & 6.06$\pm$0.04 & -3.21$\pm$0.33 & 114 & 14.36 & sdO \\
18.553344 & 52.280484 & 686415105 & 404172060533177344 & 59320$\pm$1870 & 6.05$\pm$0.06 & -2.97$\pm$0.22 & 31 & 16.44 & sdO \\
25.858836 & 32.577683 & 159006176 & 305426398708664832 & 43220$\pm$890 & 5.61$\pm$0.15 & 2.81$\pm$0.15 & 37 & 15.27 & He-sdB \\
26.668712 & 41.307165 & 698114154 & 347453684494319104 & 26660$\pm$230 & 5.08$\pm$0.03 & -0.90$\pm$0.04 & 22 & 15.83 & sdB \\
28.4816071 & 18.7996719 & 613914250 & 92226691740846080 & 31800$\pm$610 & 5.85$\pm$0.07 & -1.62$\pm$0.08 & 10 & 17.26 & sdB \\
28.953962 & 41.548932 & 631015191 & 345949758745792000 & 37560$\pm$180 & 5.69$\pm$0.03 & -3.18> & 39 & 15.74 & sdO \\
31.447815 & 40.610996 & 631006045 & 344794339527216000 & 33200$\pm$420 & 5.50$\pm$0.10 & -2.44$\pm$0.18 & 10 & 17.66 & sdB \\
31.6278862 & 54.5190312 & 380704124 & 456417279675979008 & 34470$\pm$1560 & 5.13$\pm$0.07 & -1.56$\pm$0.04 & 105 & 14.32 & sdOB \\
33.894867 & 49.427094 & 618611098 & 355574058902192768 & 37420$\pm$470 & 5.91$\pm$0.04 & -1.47$\pm$0.09 & 16 & 16.81 & sdOB \\
34.28615 & 43.681405 & 615712124 & 351536441749571200 & 35940$\pm$1350 & 5.71$\pm$0.03 & -1.50$\pm$0.04 & 29 & 15.58 & sdOB \\
35.423212 & 54.114846 & 678102180 & 455558286215251840 & 27580$\pm$330 & 5.54$\pm$0.09 & -2.78$\pm$0.54 & 13 & 16.74 & sdB \\
36.372432$^{*}$ & 28.80514 & 698702181 & 130950357400044800 & 36240$\pm$710 & 5.87$\pm$0.03 & -1.13$\pm$0.07 & 13 & 17.35 & sdOB \\
37.757164$^{*}$ & 27.718067 & 627301227 & 127674641678296704 & 47650$\pm$1400 & 5.71$\pm$0.02 & -2.78$\pm$0.29 & 79 & 15.15 & sdO \\
38.001041$^{*}$ & 33.576702 & 632206097 & 134510477267997952 & 23720$\pm$270 & 5.62$\pm$0.03 & -2.80> & 22 & 15.42 & sdB \\
42.586847 & 49.209444 & 714701194 & 438686001110484352 & 31040$\pm$160 & 5.56$\pm$0.04 & -3.01$\pm$0.79 & 26 & 15.76 & sdB \\
48.922807 & 46.869773 & 606506145 & 434851149371030272 & 39550$\pm$530 & 5.31$\pm$0.04 & -2.81$\pm$0.23 & 21 & 15.77 & sdO \\
49.744528 & 43.927658 & 616805109 & 242105008671742976 & 28450$\pm$170 & 5.94$\pm$0.02 & -3.16$\pm$0.14 & 27 & 16.51 & sdB \\
54.117831 & 46.137875 & 587807246 & 247823740446444416 & 35190$\pm$210 & 5.67$\pm$0.04 & -1.44$\pm$0.05 & 16 & 16.41 & sdOB \\
57.996529$^{*}$ & 9.640213 & 587107104 & 3302502234815943296 & 23510$\pm$170 & 5.41$\pm$0.01 & -3.00> & 35 & 15.59 & sdB \\
59.862336 & 27.08573 & 1405078 & 163565999746075264 & 33210$\pm$840 & 5.22$\pm$0.14 & -2.84$\pm$0.49 & 11 & 15.10 & sdB \\
63.957018$^{*}$ & 30.587572 & 504615117 & 165787700429000064 & 22230$\pm$320 & 5.12$\pm$0.04 & -2.64$\pm$0.16 & 13 & 16.46 & sdB \\
70.810624 & 23.217639 & 184707246 & 146588028382865280 & 36390$\pm$860 & 5.40$\pm$0.28 & -3.00> & 17 & 15.91 & sdO \\
72.124161$^{*}$ & 15.127739 & 402714066 & 3308929464395407104 & 43920$\pm$350 & 5.62$\pm$0.04 & -0.18$\pm$0.04 & 29 & 15.59 & He-sdOB \\
73.16829 & 17.529048 & 283501028 & 3406444218653682560 & 23550$\pm$270 & 5.16$\pm$0.02 & -2.39$\pm$0.08 & 19 & 16.20 & sdB \\
74.772662 & 39.631731 & 302704157 & 199210757970191744 & 32890$\pm$220 & 5.30$\pm$0.04 & -1.59$\pm$0.06 & 32 & 15.94 & sdB \\
76.710748 & 19.515218 & 202201036 & 3407876749162251648 & 50610$\pm$1040 & 5.82$\pm$0.09 & -0.21$\pm$0.12 & 33 & 16.07 & He-sdO \\
88.877241 & 61.028656 & 707916071 & 282512988705189888 & 33890$\pm$260 & 5.71$\pm$0.05 & -3.49> & 20 & 15.58 & sdB \\
88.918311$^{*}$ & 19.073818 & 330903053 & 3398598348493762944 & 63830$\pm$920 & 5.67$\pm$0.39 & -0.21$\pm$0.09 & 39 & 14.63 & He-sdO \\
89.559068 & 46.673715 & 604410009 & 197796403761616256 & 49310$\pm$870 & 5.75$\pm$0.11 & -0.16$\pm$0.06 & 21 & 17.12 & He-sdOB \\
91.999515 & 13.6144053 & 679505089 & 3344334627867111168 & 30580$\pm$1510 & 5.09$\pm$0.01 & -1.57$\pm$0.12 & 162 & 12.19 & sdB \\
92.025035 & 46.167062 & 604403022 & 963326637253435904 & 48220$\pm$480 & 5.52$\pm$0.13 & -0.14$\pm$0.12 & 18 & 17.47 & He-sdOB \\
93.23016 & 57.847462 & 679716210 & 999261490450160512 & 29270$\pm$270 & 5.43$\pm$0.03 & -2.28$\pm$0.04 & 29 & 15.81 & sdB \\
95.662806 & 46.542454 & 601111213 & 968469534172491648 & 27570$\pm$240 & 5.45$\pm$0.02 & -2.72$\pm$0.14 & 35 & 14.77 & sdB \\
97.001902 & 20.849289 & 274315180 & 3376012799112785408 & 76560$\pm$10740 & 5.10$\pm$0.21 & -0.10$\pm$0.14 & 17 & 15.81 & He-sdO \\
103.24919 & 52.713839 & 545808061 & 993265067567138432 & 31770$\pm$1220 & 5.04$\pm$0.14 & -2.35$\pm$0.19 & 17 & 15.46 & sdB \\
105.67208 & 34.633185 & 604903156 & 939579041518246272 & 34060$\pm$500 & 5.84$\pm$0.22 & -1.54$\pm$0.09 & 24 & 17.09 & sdOB \\
106.03268 & 24.199745 & 188107221 & 3380673418444759936 & 36990$\pm$1180 & 5.80$\pm$0.05 & -1.73$\pm$0.11 & 14 & 17.34 & sdOB \\
107.27312 & 22.595127 & 616616046 & 3368172319132367104 & 48480$\pm$2570 & 5.48$\pm$0.04 & -2.24$\pm$0.09 & 29 & 16.17 & sdO \\
107.72217$^{*}$ & 56.412373 & 687616080 & 988436459174352512 & 41280$\pm$540 & 5.79$\pm$0.07 & -2.87$\pm$0.51 & 50 & 14.64 & sdO \\
108.5057 & 69.55596 & 601216092 & 1109216024779190016 & 37890$\pm$460 & 5.52$\pm$0.06 & -0.15$\pm$0.05 & 14 & 16.19 & He-sdOB \\
111.25613$^{*}$ & 27.055098 & 606410248 & 872695092069071360 & 31890$\pm$420 & 5.57$\pm$0.12 & -2.29$\pm$0.06 & 34 & 16.19 & sdB \\
111.39693 & 81.847694 & 617015027 & 1142701823200960512 & 36490$\pm$140 & 5.82$\pm$0.11 & -1.48$\pm$0.07 & 29 & 15.05 & sdOB \\
112.18863 & 13.440832 & 688810176 & 3163565604772130560 & 27930$\pm$480 & 5.47$\pm$0.07 & -2.47$\pm$0.13 & 13 & 15.67 & sdB \\
112.207639 & 2.233514 & 600415096 & 3135810671409226368 & 31090$\pm$330 & 5.56$\pm$0.10 & -3.13$\pm$0.48 & 13 & 15.99 & sdB \\
112.851109 & 0.444741 & 600405172 & 3134542693985873408 & 46650$\pm$1600 & 5.74$\pm$0.25 & 1.22$\pm$0.10 & 12 & 16.31 & He-sdOB \\
113.70542 & 12.424434 & 688802020 & 3162537840574968832 & 25600$\pm$110 & 5.95$\pm$0.05 & -3.28> & 31 & 16.46 & sdB \\
113.898132 & 2.969486 & 600412059 & 3135525623021849088 & 37090$\pm$180 & 5.71$\pm$0.07 & -1.47$\pm$0.08 & 19 & 15.95 & sdOB \\
113.95052 & 26.831992 & 606408234 & 872122177791852288 & 32740$\pm$550 & 5.79$\pm$0.19 & -1.75$\pm$0.16 & 12 & 17.99 & sdB \\
114.35958 & 12.757287 & 688801151 & 3162576078665811840 & 29250$\pm$250 & 5.54$\pm$0.10 & -3.00> & 11 & 17.72 & sdB \\
118.30847$^{*}$ & 11.211171 & 605805093 & 3150707232898463616 & 29360$\pm$60 & 5.42$\pm$0.01 & -2.44$\pm$0.04 & 52 & 15.60 & sdB \\
118.3712975 & 23.4100853 & 689603199 & 675213084211549696 & 34070$\pm$40 & 5.75$\pm$0.01 & -1.68$\pm$0.03 & 122 & 13.27 & sdOB \\
123.243814$^{*}$ & 0.731455 & 641316208 & 3089571878131969792 & 28390$\pm$340 & 5.42$\pm$0.03 & -2.61$\pm$0.13 & 24 & 14.59 & sdB \\
124.735819 & 39.901597 & 642010128 & 909317797165729024 & 22910$\pm$210 & 5.44$\pm$0.07 & -3.48$\pm$0.29 & 38 & 14.94 & sdB \\
\hline\noalign{\smallskip} 
  \end{tabularx}
  \tablenotetext{a}{Stars labeled with $\ast$ also appear in the hot subdwarf catalog of Geier et al. (2017).} 
 \tablenotetext{b}{"$>$" denotes an upper limit of $\mathrm{log}(n\mathrm{He}/n\mathrm{H})$ for the object.} 
\end{table*}

\setcounter{table}{0} 
\begin{table*}
\tiny

 \begin{minipage}{160mm}
  \caption{Continued.}
  \end{minipage}\\
  \centering
    \begin{tabularx}{16.0cm}{lllccccccccccccccX}
\hline\noalign{\smallskip}
RA\tablenotemark{a}  & DEC   &obsid & source\_id 
& $T_\mathrm{eff}$ &  $\mathrm{log}\ g$ & $\mathrm{log}(n\mathrm{He}/n\mathrm{H})$\tablenotemark{b}
&SNRU & $G$  & spclass  \\
 LAMOST &  LAMOST& LAMOST&Gaia  &(K)&($\mathrm{cm\ s^{-2}}$)& && Gaia(mag)& \\
\hline\noalign{\smallskip}
124.998525$^{*}$ & 22.6836111 & 602216224 & 676607952150024448 & 31300$\pm$180 & 5.69$\pm$0.07 & -1.79$\pm$0.07 & 17 & 15.64 & sdB \\
125.2234083$^{*}$ & 0.1455028 & 641315139 & 3077510098136276480 & 29030$\pm$90 & 5.65$\pm$0.02 & -1.96$\pm$0.16 & 31 & 15.18 & sdB \\
126.190367$^{*}$ & 23.255656 & 602216150 & 678116344664890368 & 29900$\pm$220 & 5.47$\pm$0.02 & -2.95$\pm$0.14 & 38 & 15.34 & sdB \\
126.28563$^{*}$ & 48.675328 & 615105036 & 930960515328049536 & 29760$\pm$300 & 5.62$\pm$0.07 & -2.96$\pm$0.33 & 17 & 16.98 & sdB \\
129.0820292$^{*}$ & 20.9636028 & 699412249 & 664631178147534720 & 31370$\pm$440 & 5.52$\pm$0.11 & -2.44$\pm$0.15 & 10 & 16.34 & sdB \\
130.255486$^{*}$ & 39.938389 & 642006098 & 911573758803336960 & 29380$\pm$140 & 5.69$\pm$0.03 & -2.41$\pm$0.06 & 31 & 15.45 & sdB \\
131.034178$^{*}$ & 31.03639 & 130107157 & 706479277895031040 & 29320$\pm$340 & 5.39$\pm$0.03 & -2.14$\pm$0.07 & 30 & 14.56 & sdB \\
131.19586$^{*}$ & 11.652792 & 420803011 & 601862464498177664 & 28580$\pm$190 & 5.37$\pm$0.01 & -2.53$\pm$0.07 & 38 & 16.13 & sdB \\
134.47774$^{*}$ & 38.314391 & 711713027 & 719606175420853888 & 30870$\pm$230 & 5.52$\pm$0.09 & -2.62> & 14 & 15.68 & sdB \\
137.90359$^{*}$ & 27.877858 & 186807004 & 698115121143554176 & 46750$\pm$1780 & 5.72$\pm$0.03 & -2.87$\pm$0.21 & 17 & 17.00 & sdO \\
143.8201083$^{*}$ & 22.8279833 & 606605119 & 644079931432984704 & 37000$\pm$430 & 5.64$\pm$0.07 & -2.05$\pm$0.07 & 38 & 16.29 & sdOB \\
144.90796 & 17.664899 & 606102183 & 620899404525808768 & 79240$\pm$5590 & 6.58$\pm$0.06 & -2.19$\pm$0.36 & 17 & 17.48 & sdO \\
147.75537$^{*}$ & 3.7991754 & 709312041 & 3849462024992532608 & 29500$\pm$110 & 5.41$\pm$0.04 & -2.63$\pm$0.62 & 19 & 15.89 & sdB \\
148.3206292$^{*}$ & 15.5617194 & 731215192 & 616743220508208896 & 41010$\pm$10 & 5.66$\pm$0.05 & 1.80$\pm$0.07 & 35 & 15.52 & He-sdOB \\
150.4163042 & -3.0035611 & 723303210 & 3829267569803099776 & 31220$\pm$470 & 5.47$\pm$0.07 & -2.81$\pm$0.53 & 16 & 16.71 & sdB \\
160.469575$^{*}$ & 21.675766 & 712413228 & 3987913113277693184 & 33410$\pm$90 & 5.75$\pm$0.02 & -2.17$\pm$0.04 & 48 & 13.07 & sdOB \\
162.3896667$^{*}$ & 18.7115278 & 215810196 & 3983291213071411712 & 29500$\pm$340 & 5.14$\pm$0.06 & -2.44$\pm$0.07 & 34 & 14.92 & sdB \\
178.016827$^{*}$ & 39.140844 & 657402229 & 4034502959999559168 & 55800$\pm$2480 & 5.48$\pm$0.09 & -2.03$\pm$0.17 & 16 & 15.36 & sdO \\
195.10638$^{*}$ & 0.7583765 & 144103116 & 3689536684343245312 & 37940$\pm$730 & 6.02$\pm$0.14 & -1.42$\pm$0.09 & 16 & 15.72 & sdOB \\
199.69624$^{*}$ & 44.595021 & 739312049 & 1550490241899314560 & 42850$\pm$100 & 5.73$\pm$0.01 & 0.92$\pm$0.07 & 68 & 14.77 & He-sdOB \\
204.2248833$^{*}$ & 11.4347944 & 734713132 & 3738606616980353664 & 37850$\pm$260 & 5.87$\pm$0.06 & -1.50$\pm$0.05 & 14 & 16.34 & sdOB \\
204.54297$^{*}$ & 43.295307 & 449115138 & 1501713500909166208 & 34600$\pm$560 & 5.19$\pm$0.06 & -1.12$\pm$0.15 & 12 & 16.77 & sdOB \\
206.58844$^{*}$ & 22.810201 & 660604235 & 1251408094001504640 & 35240$\pm$2100 & 5.94$\pm$0.17 & -0.16$\pm$0.08 & 14 & 17.15 & He-sdOB \\
206.7520875$^{*}$ & 11.1901194 & 733615158 & 3727881843124118400 & 23510$\pm$40 & 5.60$\pm$0.01 & -3.00> & 63 & 14.96 & sdB \\
208.76946$^{*}$ & -2.506063 & 651513250 & 3657799934042253952 & 45740$\pm$1000 & 5.65$\pm$0.04 & -1.80$\pm$0.13 & 35 & 12.06 & sdO \\
211.43857$^{*}$ & 1.7386288 & 732404097 & 3661331668469980416 & 27770$\pm$90 & 5.27$\pm$0.02 & -2.02$\pm$0.05 & 37 & 15.81 & sdB \\
212.732694$^{*}$ & 9.548705 & 723502076 & 3723006814724972416 & 36840$\pm$220 & 5.83$\pm$0.02 & -1.66$\pm$0.06 & 60 & 14.05 & sdOB \\
213.954395$^{*}$ & 11.2038595 & 723504163 & 1225417739360402048 & 41500$\pm$100 & 5.50$\pm$0.11 & 1.67$\pm$0.25 & 22 & 16.03 & He-sdOB \\
221.29285 & 14.229163 & 343616178 & 1185738013981539840 & 57000$\pm$3130 & 6.52$\pm$0.07 & -2.61> & 17 & 16.39 & sdO \\
221.3759375$^{*}$ & 17.4645 & 657010071 & 1234828283288291840 & 71170$\pm$6170 & 6.90$\pm$0.07 & -2.08$\pm$0.24 & 28 & 16.23 & sdO \\
223.01644$^{*}$ & 45.558239 & 742605036 & 1586890398971315200 & 49470$\pm$1360 & 5.65$\pm$0.08 & -1.98$\pm$0.14 & 22 & 17.23 & sdO \\
224.526717 & 8.858398 & 651102213 & 1161864283648012160 & 22500$\pm$160 & 5.48$\pm$0.03 & -3.22$\pm$0.10 & 54 & 14.64 & sdB \\
224.5663917$^{*}$ & 37.0047194 & 743709156 & 1295107633891682944 & 49610$\pm$600 & 6.04$\pm$0.10 & 0.00$\pm$0.04 & 22 & 17.30 & He-sdOB \\
224.8688067$^{*}$ & 19.0638675 & 657013168 & 1188933362275187200 & 36420$\pm$540 & 6.00$\pm$0.03 & -1.53$\pm$0.04 & 44 & 14.25 & sdOB \\
227.154254$^{*}$ & 10.053918 & 651108142 & 1167834597427267456 & 35880$\pm$290 & 5.79$\pm$0.11 & -1.13$\pm$0.12 & 47 & 15.10 & sdOB \\
231.78$^{*}$ & 10.270154 & 565710176 & 1165815825359631232 & 33230$\pm$730 & 5.15$\pm$0.05 & -1.97$\pm$0.08 & 13 & 16.14 & sdB \\
234.67853$^{*}$ & 9.5784135 & 565707040 & 1165071009310870912 & 35840$\pm$110 & 5.62$\pm$0.03 & -0.93$\pm$0.04 & 19 & 15.73 & sdOB \\
239.4949003$^{*}$ & 14.0390339 & 740903219 & 1191689807863866112 & 30520$\pm$310 & 5.85$\pm$0.03 & -2.88$\pm$0.67 & 33 & 15.37 & sdB \\
239.8282$^{*}$ & 5.6004099 & 744014072 & 4426623509802852352 & 30650$\pm$170 & 5.53$\pm$0.03 & -2.96$\pm$0.18 & 16 & 16.88 & sdB \\
241.1203583$^{*}$ & 14.8469639 & 740909091 & 1192038902805203328 & 32460$\pm$90 & 5.88$\pm$0.06 & -3.06$\pm$0.47 & 28 & 16.04 & sdB \\
243.18797$^{*}$ & 4.2115442 & 744007193 & 4437254653372798848 & 45700$\pm$350 & 5.65$\pm$0.05 & 0.96$\pm$0.17 & 22 & 16.03 & He-sdOB \\
245.7361$^{*}$ & 47.514196 & 743008201 & 1410860511508492288 & 28100$\pm$220 & 5.66$\pm$0.05 & -1.79$\pm$0.05 & 24 & 16.24 & sdB \\
247.4703492$^{*}$ & 11.0840364 & 663715062 & 4458994472154612480 & 27990$\pm$160 & 5.42$\pm$0.00 & -2.61$\pm$0.04 & 33 & 14.35 & sdB \\
247.9622125$^{*}$ & 48.0752639 & 743006102 & 1410554774260311808 & 38780$\pm$420 & 5.55$\pm$0.06 & -0.48$\pm$0.09 & 11 & 17.12 & He-sdOB \\
250.878644$^{*}$ & 51.415874 & 585102152 & 1413338325384928128 & 35940$\pm$550 & 5.09$\pm$0.03 & -2.02$\pm$0.12 & 20 & 16.17 & sdOB \\
251.64371$^{*}$ & 26.6312 & 743504173 & 1307252843628956672 & 39810$\pm$490 & 6.27$\pm$0.04 & 2.20$\pm$0.12 & 35 & 16.14 & He-sdB \\
254.756648$^{*}$ & 29.042889 & 739510211 & 1309437641952913920 & 27720$\pm$700 & 5.61$\pm$0.12 & -2.89$\pm$0.19 & 12 & 16.11 & sdB \\
254.990298$^{*}$ & 28.848331 & 739510216 & 1308678016856993920 & 37520$\pm$660 & 5.73$\pm$0.00 & -3.18$\pm$0.19 & 96 & 14.39 & sdO \\
255.76656 & 15.138432 & 745914113 & 4545907052398514432 & 27460$\pm$400 & 5.34$\pm$0.04 & -1.88$\pm$0.77 & 17 & 17.12 & sdB \\
257.76869 & 11.765573 & 664314127 & 4540919083539644672 & 29410$\pm$410 & 5.61$\pm$0.05 & -2.53$\pm$0.14 & 10 & 17.80 & sdB \\
258.2631$^{*}$ & 16.178565 & 745911237 & 4546882216133354752 & 35550$\pm$420 & 5.82$\pm$0.03 & -1.61$\pm$0.04 & 40 & 16.27 & sdOB \\
259.33566 & 9.6920351 & 664305033 & 4491582966009639040 & 54240$\pm$1590 & 5.25$\pm$0.20 & 1.76$\pm$0.21 & 26 & 16.94 & He-sdOB \\
259.824376$^{*}$ & 47.372495 & 745114017 & 1365071418489267584 & 38620$\pm$530 & 5.87$\pm$0.06 & -2.58$\pm$0.25 & 13 & 15.79 & sdOB \\
260.03809 & 15.843371 & 745912208 & 4546952309999524096 & 42240$\pm$590 & 5.13$\pm$0.05 & 2.10$\pm$0.01 & 27 & 16.43 & He-sdOB \\
262.26331 & 17.326944 & 742209223 & 4550175420961718528 & 37750$\pm$660 & 5.17$\pm$0.08 & -2.65$\pm$0.19 & 10 & 16.31 & sdB \\
262.8484903 & 46.2253286 & 566204175 & 1361931728676649984 & 28770$\pm$1270 & 5.65$\pm$0.07 & -1.98$\pm$0.10 & 22 & 17.15 & sdB \\
264.4181 & 19.372917 & 742808226 & 4550885847209172736 & 32300$\pm$110 & 5.69$\pm$0.03 & -2.13$\pm$0.10 & 15 & 16.99 & sdOB \\
270.030251 & 31.577103 & 663811042 & 4603104815507642752 & 26970$\pm$440 & 5.52$\pm$0.05 & -3.23$\pm$0.24 & 27 & 15.23 & sdB \\
271.34282 & 15.200315 & 746410046 & 4498502433203434368 & 28260$\pm$100 & 5.30$\pm$0.02 & -3.07$\pm$0.15 & 14 & 16.92 & sdB \\
271.95262 & 14.507648 & 746402150 & 4498196768968750336 & 28730$\pm$290 & 5.39$\pm$0.04 & -3.00$\pm$0.18 & 38 & 15.91 & sdB \\
272.36637 & 16.769945 & 746403210 & 4502091509736932608 & 65660$\pm$11140 & 5.68$\pm$0.06 & -1.79$\pm$0.15 & 15 & 17.77 & sdO \\
272.76169 & 17.633307 & 746415156 & 4526224282435916544 & 44410$\pm$1260 & 5.97$\pm$0.19 & 0.54$\pm$0.12 & 19 & 17.10 & He-sdOB \\
273.0417 & 18.131682 & 746411012 & 4526347084141410432 & 30170$\pm$50 & 5.41$\pm$0.01 & -3.08$\pm$0.84 & 48 & 15.48 & sdB \\
273.04685 & 17.927912 & 746411078 & 4526236102186660352 & 23640$\pm$420 & 5.03$\pm$0.07 & -2.30$\pm$0.15 & 12 & 17.40 & sdB \\
273.32165 & 34.316932 & 743615051 & 4605158393990687360 & 31140$\pm$50 & 5.37$\pm$0.04 & -1.88$\pm$0.29 & 23 & 17.15 & sdB \\
274.13538 & 34.930314 & 743615206 & 4605575383775753856 & 47780$\pm$1720 & 5.48$\pm$0.03 & -2.92$\pm$0.11 & 59 & 16.02 & sdO \\
274.83431 & 18.178403 & 746412176 & 4523666131196149632 & 27630$\pm$220 & 5.14$\pm$0.03 & -2.95$\pm$0.21 & 13 & 16.44 & sdB \\
274.88444 & 6.0783654 & 742405150 & 4476966603891088640 & 36240$\pm$580 & 5.37$\pm$0.16 & -3.00> & 13 & 17.13 & sdB \\
274.95133$^{*}$ & 33.369344 & 743604247 & 4592825172063276288 & 28380$\pm$180 & 5.44$\pm$0.02 & -2.55$\pm$0.07 & 32 & 16.44 & sdB \\
275.51348 & 10.739026 & 746714189 & 4483659679067140992 & 28940$\pm$390 & 5.05$\pm$0.07 & -1.44$\pm$0.07 & 12 & 17.37 & sdB \\
288.82889$^{*}$ & 42.93705 & 664703151 & 2102745688098547840 & 38420$\pm$960 & 5.54$\pm$0.16 & -3.05$\pm$0.29 & 25 & 14.49 & sdO \\
291.81268$^{*}$ & 38.45518 & 664011152 & 2052684550030830464 & 39890$\pm$50 & 5.38$\pm$0.15 & 0.54$\pm$0.04 & 32 & 15.61 & He-sdOB \\
292.78702$^{*}$ & 43.416039 & 664613110 & 2125895669204184832 & 66880$\pm$11150 & 5.15$\pm$0.03 & -1.01$\pm$0.35 & 43 & 13.59 & sdO \\
293.53371 & 35.000895 & 664007120 & 2048109069842534912 & 34560$\pm$810 & 5.71$\pm$0.06 & -1.41$\pm$0.06 & 13 & 15.34 & sdOB \\
293.86659 & 35.732989 & 664008145 & 2048434490916786176 & 29610$\pm$300 & 5.63$\pm$0.09 & -2.62$\pm$0.20 & 16 & 15.77 & sdB \\
303.0673667$^{*}$ & 8.2691222 & 587214212 & 4251149700348007680 & 27730$\pm$220 & 5.39$\pm$0.06 & -2.98$\pm$0.12 & 49 & 14.62 & sdB \\
303.406715 & 9.467058 & 746301022 & 4299431347569705216 & 33430$\pm$110 & 5.15$\pm$0.02 & -2.62$\pm$0.06 & 119 & 12.41 & sdOB \\
305.47392 & 6.488096 & 587308185 & 4249752113691558144 & 30820$\pm$580 & 5.65$\pm$0.07 & -2.85> & 12 & 17.44 & sdB \\
305.67635 & 7.2876099 & 587304110 & 4249937660575808000 & 29270$\pm$130 & 5.42$\pm$0.03 & -3.10$\pm$0.14 & 17 & 17.07 & sdB \\
\hline\noalign{\smallskip} 
  \end{tabularx}
\end{table*}

\setcounter{table}{0}
\begin{table*}
\tiny

 \begin{minipage}{160mm}
  \caption{Continued.}
  \end{minipage}\\
  \centering
    \begin{tabularx}{16.0cm}{lllccccccccccccccX}
\hline\noalign{\smallskip}
RA\tablenotemark{a}  & DEC   &obsid & source\_id 
& $T_\mathrm{eff}$ &  $\mathrm{log}\ g$ & $\mathrm{log}(n\mathrm{He}/n\mathrm{H})$\tablenotemark{b}
&SNRU & $G$  & spclass  \\
 LAMOST &  LAMOST& LAMOST&Gaia  &(K)&($\mathrm{cm\ s^{-2}}$)& && Gaia(mag)& \\
\hline\noalign{\smallskip}
317.36219 & 37.139663 & 680412153 & 1868767831308190976 & 50100$\pm$1630 & 5.53$\pm$0.04 & -1.68$\pm$0.06 & 41 & 15.28 & sdO \\
317.503089 & 15.486887 & 677702178 & 1760662130066900352 & 43540$\pm$520 & 5.45$\pm$0.11 & 1.53$\pm$0.50 & 18 & 16.40 & He-sdOB \\
318.85231 & 38.577478 & 593803082 & 1965019835117424000 & 27670$\pm$380 & 5.34$\pm$0.05 & -2.59$\pm$0.30 & 12 & 17.63 & sdB \\
318.88115 & 12.665982 & 592402156 & 1746789866736764800 & 29120$\pm$370 & 5.55$\pm$0.05 & -2.65$\pm$0.16 & 30 & 16.02 & sdB \\
319.51366$^{*}$ & 14.681637 & 592414174 & 1759463868550744576 & 28960$\pm$70 & 5.60$\pm$0.02 & -3.00> & 62 & 15.06 & sdB \\
320.52329 & 21.686536 & 593111200 & 1790728889707996800 & 27620$\pm$180 & 5.50$\pm$0.02 & -2.61$\pm$0.03 & 28 & 15.17 & sdB \\
320.87692$^{*}$ & 0.710801 & 254804012 & 2690967057290240000 & 35050$\pm$280 & 5.89$\pm$0.04 & -0.78$\pm$0.04 & 24 & 16.87 & He-sdOB \\
320.98619 & 15.55655 & 592415028 & 1783640205099886336 & 49100$\pm$2030 & 5.54$\pm$0.20 & 0.60$\pm$0.08 & 18 & 16.71 & He-sdOB \\
321.5339 & 2.759411 & 677911144 & 2691867011851791744 & 29550$\pm$370 & 5.90$\pm$0.07 & -2.58$\pm$0.20 & 12 & 16.16 & sdB \\
321.70608 & 15.760201 & 592412140 & 1783628003096502144 & 34010$\pm$470 & 5.84$\pm$0.06 & -1.61$\pm$0.07 & 16 & 17.05 & sdOB \\
321.797821$^{*}$ & 0.196107 & 677904113 & 2687870218366060416 & 29570$\pm$90 & 5.52$\pm$0.03 & -2.98$\pm$0.08 & 61 & 14.57 & sdB \\
322.80103$^{*}$ & 11.493389 & 679903086 & 1745849337621677184 & 37210$\pm$290 & 5.89$\pm$0.02 & -1.66$\pm$0.04 & 55 & 15.94 & sdOB \\
323.6435$^{*}$ & 9.6801009 & 679901047 & 1741581170917641728 & 36590$\pm$140 & 5.79$\pm$0.05 & -1.44$\pm$0.04 & 49 & 15.55 & sdOB \\
335.57083 & 26.93794 & 594102069 & 1881671180067646080 & 28760$\pm$270 & 5.38$\pm$0.08 & -3.41$\pm$0.45 & 26 & 16.77 & sdB \\
335.57458 & 27.588819 & 594105250 & 1881776668761026688 & 50650$\pm$2760 & 5.56$\pm$0.23 & -2.30$\pm$0.17 & 50 & 15.50 & sdO \\
340.48328 & 17.803049 & 601803112 & 2832879034517348608 & 30510$\pm$600 & 5.46$\pm$0.06 & -3.26> & 18 & 17.53 & sdB \\
341.26507$^{*}$ & 32.364203 & 680503235 & 1890677009230168704 & 31020$\pm$240 & 5.70$\pm$0.04 & -2.74$\pm$0.12 & 89 & 14.00 & sdB \\
341.89198$^{*}$ & 33.011002 & 680503175 & 1890817059523265024 & 26720$\pm$230 & 5.55$\pm$0.04 & -2.82$\pm$0.09 & 33 & 16.00 & sdB \\
344.17032$^{*}$ & 29.762963 & 606014218 & 1886200725594365568 & 36060$\pm$1050 & 6.01$\pm$0.07 & -1.48$\pm$0.08 & 34 & 16.35 & sdOB \\
344.54285$^{*}$ & 40.727771 & 604109042 & 1930945626165804032 & 43920$\pm$290 & 5.55$\pm$0.03 & 0.85$\pm$0.08 & 26 & 15.99 & He-sdOB \\
346.28713 & 30.454833 & 677603238 & 1886482196274463488 & 35340$\pm$860 & 5.88$\pm$0.12 & -1.48$\pm$0.10 & 12 & 17.18 & sdOB \\
347.33042 & 36.899804 & 678214001 & 1915148289773812352 & 33520$\pm$790 & 5.68$\pm$0.04 & -1.41$\pm$0.05 & 13 & 16.41 & sdOB \\
350.23543 & 34.39527 & 678205023 & 1912906626082386304 & 36100$\pm$140 & 5.79$\pm$0.02 & -1.40$\pm$0.03 & 44 & 15.43 & sdOB \\
350.90066 & 45.217675 & 587411151 & 1937879932466496896 & 23950$\pm$1700 & 5.81$\pm$0.11 & -2.72$\pm$0.22 & 10 & 17.07 & sdB \\
351.37223 & 41.518334 & 587405077 & 1923590271333313792 & 38370$\pm$660 & 5.90$\pm$0.07 & -1.41$\pm$0.07 & 31 & 16.30 & sdOB \\
352.05393 & 29.892664 & 593502206 & 2869717686274246400 & 45030$\pm$2480 & 5.55$\pm$0.54 & 1.19$\pm$0.17 & 14 & 16.71 & He-sdOB \\
352.23596 & 49.468338 & 689005189 & 1942912367126946304 & 40640$\pm$30 & 5.59$\pm$0.03 & 1.88$\pm$0.06 & 19 & 17.38 & He-sdB \\
352.34432 & 32.233162 & 593503117 & 2872454748672529280 & 31070$\pm$280 & 5.49$\pm$0.07 & -2.59$\pm$0.26 & 22 & 16.93 & sdB \\
353.68925 & 51.004629 & 689015198 & 1944738965178800384 & 37140$\pm$490 & 5.53$\pm$0.03 & -3.23> & 45 & 16.03 & sdO \\
354.17191 & 31.533466 & 593504113 & 2871378846483069184 & 33470$\pm$280 & 5.96$\pm$0.04 & -2.46$\pm$0.12 & 24 & 16.86 & sdB \\
354.84683 & 46.912366 & 678715200 & 1939195223251633024 & 30560$\pm$750 & 5.82$\pm$0.08 & -1.53$\pm$0.06 & 16 & 17.78 & sdB \\
355.66937$^{*}$ & 43.91102 & 678701017 & 1925782766239946624 & 39780$\pm$480 & 5.29$\pm$0.05 & -2.75$\pm$0.14 & 33 & 16.15 & sdO \\
358.62099 & 35.560802 & 602605007 & 2878501890826543616 & 32860$\pm$940 & 5.43$\pm$0.09 & -1.69$\pm$0.09 & 16 & 17.43 & sdOB \\
\hline\noalign{\smallskip} 
  \end{tabularx}
\end{table*}

\begin{figure}
\centering
\begin{minipage}[c]{0.42\textwidth}
\includegraphics [width=75mm]{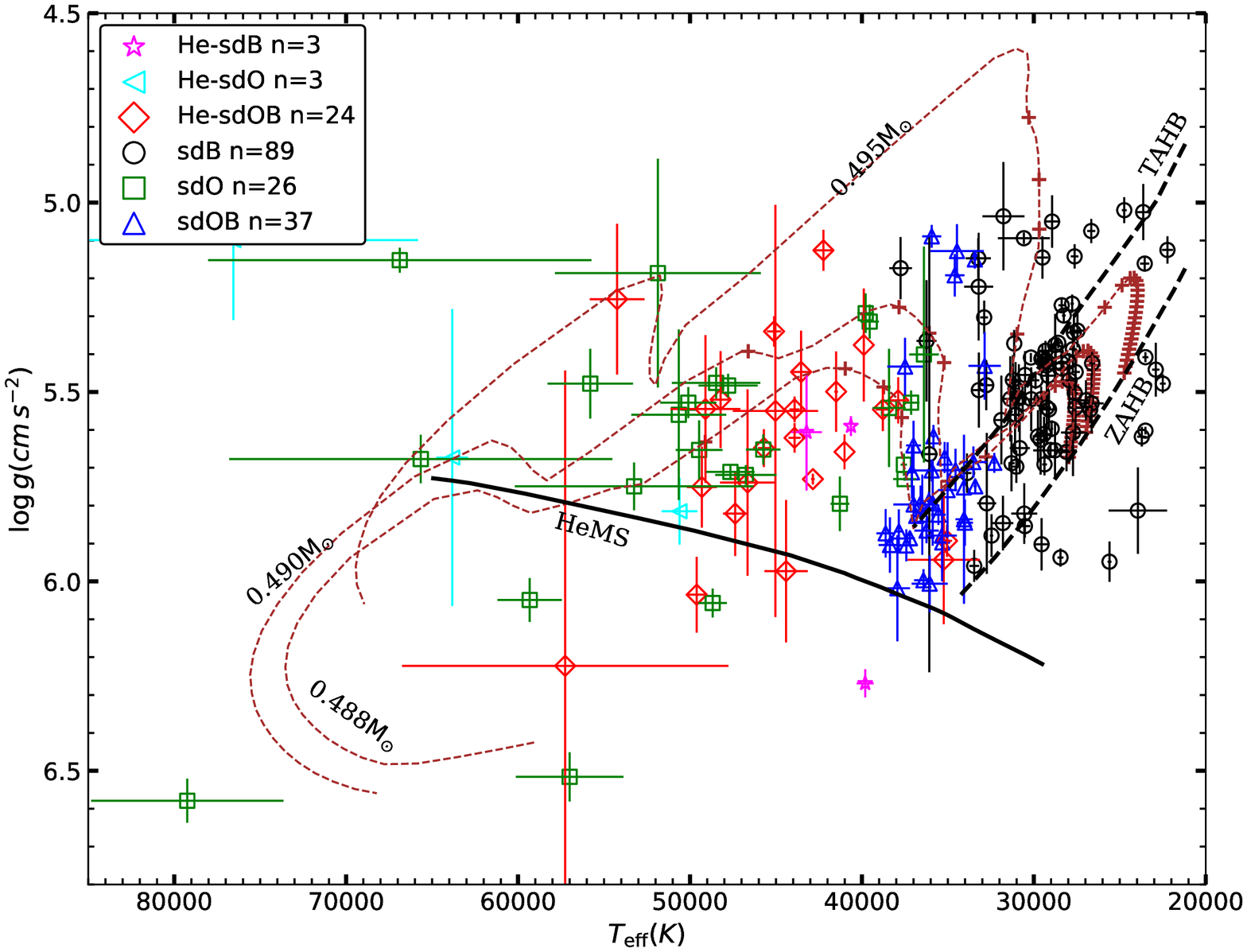}
\centerline{ a }
\end{minipage}%
\begin{minipage}[c]{0.42\textwidth}
\includegraphics [width=75mm]{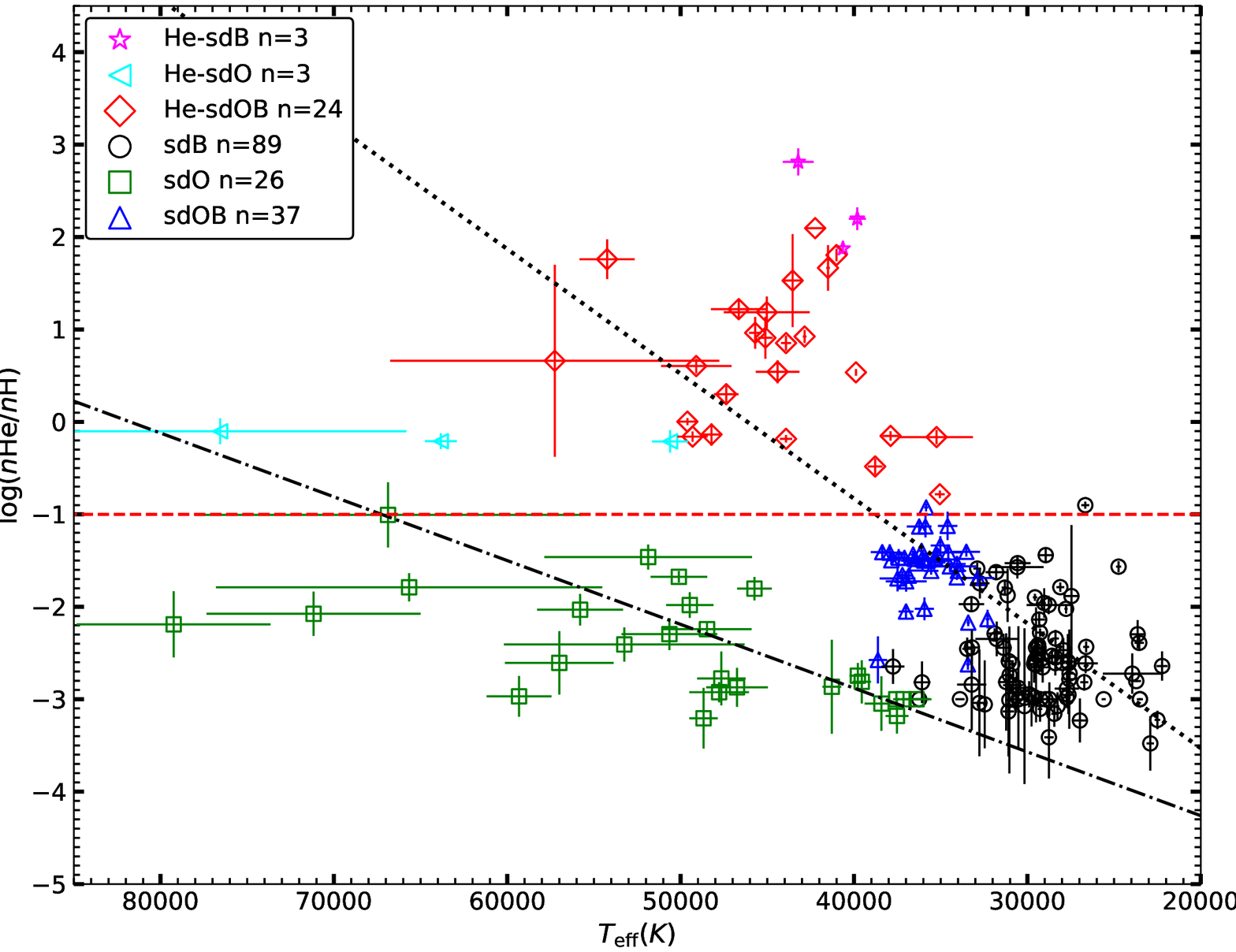} 
\centerline{ b }
\end{minipage}\\%
\begin{minipage}[c]{0.45\textwidth}
\includegraphics [width=75mm]{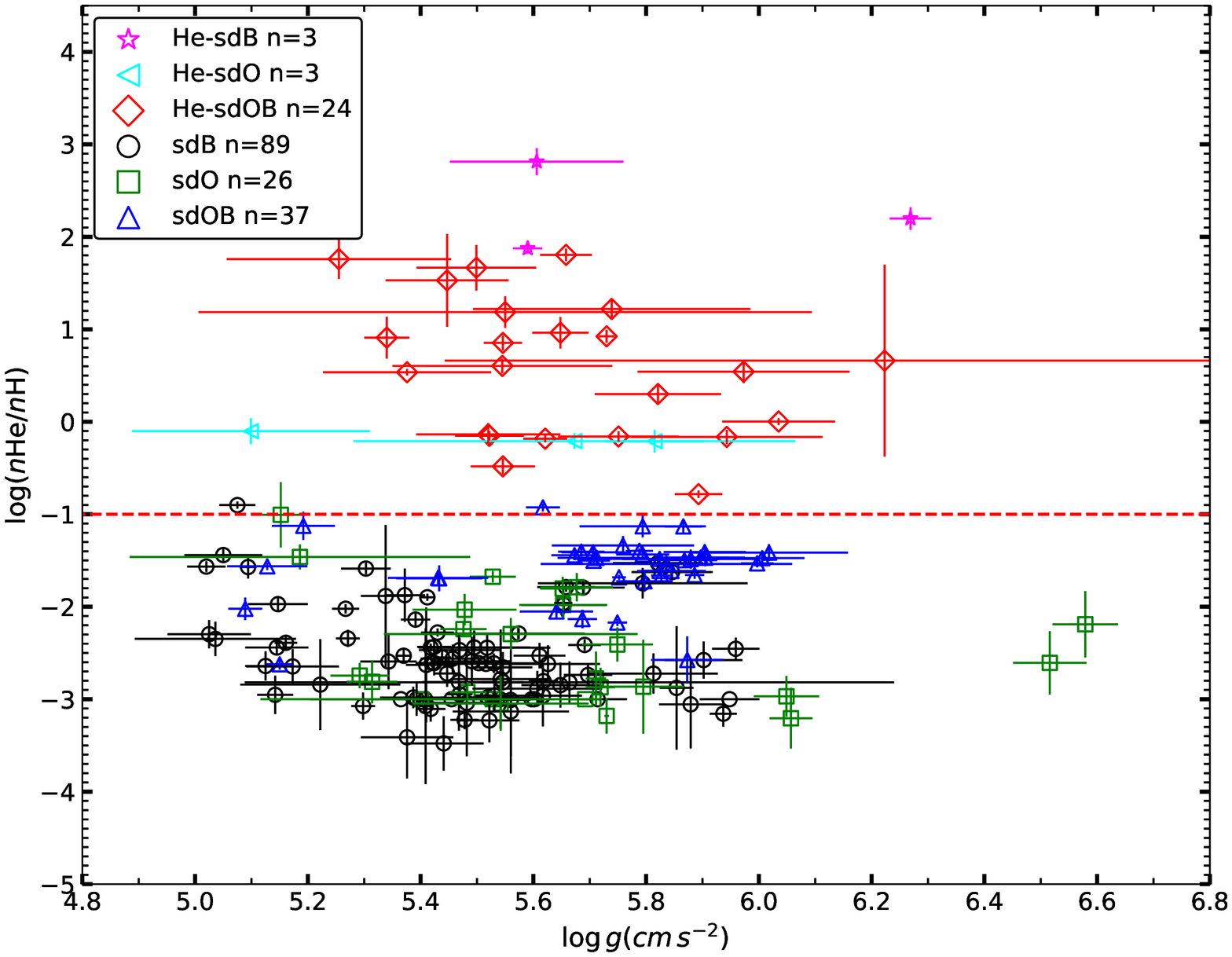} 
\centerline{ c }
\end{minipage}%

\caption{Atmospheric parameter diagrams for the 182 hot 
subdwarf stars identified in this study. 
The markers and number counts for different types of hot subdwarfs 
are showed in the upper-left box of each panel.  
Panel a: $T_{\rm eff}$-$\log{g}$ diagram. The zero-age horizontal branch (ZAHB) 
and terminal-age horizontal branch (TAHB) sequences 
with [Fe/H]= -1.48 from Dorman et al. (1993) are denoted by dashed lines. 
The He-MS from Paczy\'nski (1971) is marked by the black solid line. 
Three evolution tracks for hot HB stars from Dorman et al. (1993) 
are showed with brown dotted curves, 
and their masses from top to bottom are 0.495, 0.490 and 0.488 $M_{\odot}$, respectively. 
Panel b: $T_{\rm eff}$-$\mathrm{log}(n\mathrm{He}/n\mathrm{H})$ 
diagram. The black dotted line and dot-dashed line are the linear regression lines fitted  
by Edelmann et al. (2003) and N\'emeth et al. (2012), respectively.  
Panel c: $\log{g}$-$\mathrm{log}(n\mathrm{He}/n\mathrm{H})$ diagram. 
The red horizontal dashed line in panel b and c denotes the solar value of the 
He abundance (e.g., $\mathrm{log}(n\mathrm{He}/n\mathrm{H})$ = -1).}  
\end{figure} 

Fig 3 shows the parameter diagrams for the 182 single-lined hot subdwarf stars. 
In panel a, the majority of sdB stars (black circles) are in a region 
that is well defined by the ZAHB and TAHB (e.g., centered at $T_{\rm eff}$ = 28000 K and 
$\log{g}$ = 5.5 $\mathrm{cm\ s^{-2}}$), 
which demonstrates that these stars 
are undergoing helium burning in their cores. On the other hand, 
sdOB stars (blue up triangles) which cluster around at 
$T_{\rm eff}$ = 34000 K and $\log{g}$ = 5.8 $\mathrm{cm\ s^{-2}}$, 
present higher effective temperatures and   
$\log{g}$ than sdB stars. 
SdO stars (green squares) and  He-sdO stars (aqua left triangles)  
present the highest effective temperatures in our sample, e.g., 
most of them have $T_{\rm eff}>$ 40000 K, but with a wide range of 
$\log{g}$. He-sdOB stars (red diamonds) cluster around at 
$T_{\rm eff}$ = 45000 K and $\log{g}$ = 5.6 $\mathrm{cm\ s^{-2}}$. 
The 2 He-sdB stars (magenta stars), which present the highest 
He abundance in our sample, are located in the area very close 
to our He-sdOB stars in panel a. 
The hot subdwarf samples share similar characteristics in the $T_{\rm eff}$-$\log{g}$ diagram  
with our previous study (e.g., see panel a of Fig 4 in Lei et al. 2019b 
and Fig 6 in Lei et al. 2018).  

Panel b in Fig 3 shows the $T_{\rm eff}$-$\mathrm{log}(n\mathrm{He}/n\mathrm{H})$ diagram 
for our hot subdwarf sample. Two distinct helium sequences, e.g., a He-rich sequence (fitted by dotted line) and a  
He-weak sequence (fitted by dot-dashed line), 
which were discovered by Edelmann et al. (2003) and confirmed by later  studies 
(N\'emeth et al. 2012; Geier et al. 2013; Luo et al. 2016b; 
Lei et al. 2018, 2019b), are clearly present in this panel. 
As  found by Lei et al. (2019b), 
the He-rich sequence consists of sdB, sdOB, He-sdOB and He-sdB stars, while 
the He-weak sequence consists purely of sdO stars. 
Furthermore, the 3 He-sdO stars (aqua left triangles) identified in 
this study are located between the two He sequences. 
However, the physical mechanism responsible for the two He sequences of 
hot subdwarf stars is still unclear, and additional scenarios are needed. 

In panel b, one also can find a gap 
(e.g., $T_{\rm eff}$ = 40000 K and $\mathrm{log}(n\mathrm{He}/n\mathrm{H})$ = 0.0)  
in He-sdOB stars (red diamonds), which splits the He-sdOB 
stars into two subgroups, e.g., a subgroup with higher He abundances and temperatures, and 
the other subgroup with lower He abundance and temperatures. 
With larger size of He-sdOB stars, this gap is more clearly present in 
panel b of Fig 4 in Lei et al. (2019b). As  
discussed in Lei et al. (2019b), the gap also 
appears in the EHB stars of GC $\omega$ Cen, but 
the star fraction of the two subgroups between field 
hot subdwarf stars and $\omega$ Cen EHB stars are very different. 
Moreover, 3 He-sdB stars are found in this study, 
which present the highest He abundances (e.g., $\mathrm{log}(n\mathrm{He}/n\mathrm{H})>$ 2.0) 
in our sample. However, the most He-rich stars are completely missing from the 
EHB stars of $\omega$ Cen (see Fig 6 in Lei et al. 2019b and the text therein 
for detailed discussion). All these results point towards a  different 
formation of field hot subdwarf stars and GC EHB stars, 
and provide a strict observational limit on the evolution models for 
the two types of objects.

\begin{figure} 
\centering
\includegraphics [width=160mm]{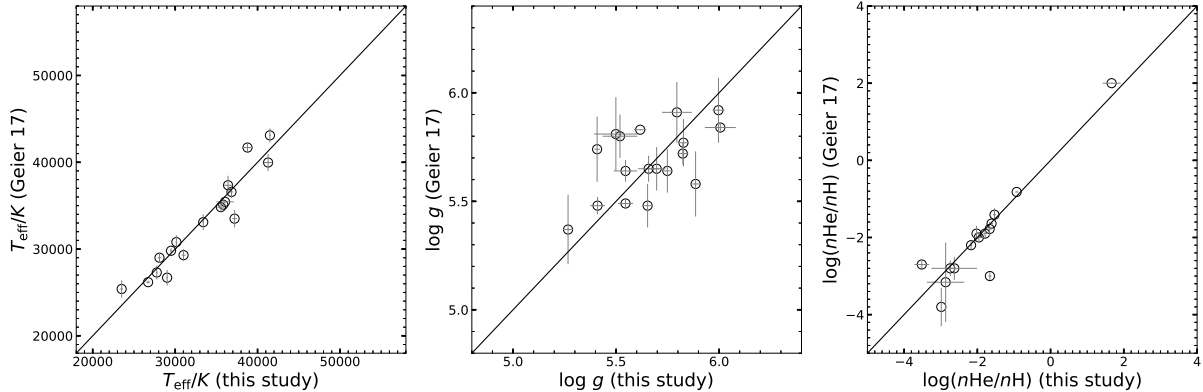}
\caption{Atmospheric parameters comparison with the catalog of Geier et al. (2017) 
for the common stars.  }
\end{figure}  

As described above, we have 74  stars in common with the 
hot subdwarf catalog of Geier et al. (2017), therefore, we 
compared the parameters of the stars obtained in 
this study and the parameters reported in the catalog of Geier et al.(2017) 
as long as their atmospheric parameters are available. Fig 4 presents the 
results from this comparison. Horizontal coordinates denote the 
parameter values obtained in this study, while vertical coordinates 
represent the values from Geier et al. (2017). 
As we see in Fig 4, the values of $T_{\rm eff}$ (e.g., left panel) 
and $\mathrm{log}(n\mathrm{He}/n\mathrm{H})$ (e.g., right panel) obtained 
in this study  are well consistent with the ones reported in 
Geier et al. (2017). Although the comparison of $\log{g}$ (middle panel) 
shows a little larger dispersion than the other two parameters  
(e.g., $T_{\rm eff}$ and $\mathrm{log}(n\mathrm{He}/n\mathrm{H})$), 
the values are still comparable when 
the large systematic errors that affect $\log{g}$ are considered. 
One source of these errors stem from the different implementations of 
Stark broadening tables in various model atmosphere codes. Another 
source is the variable observational data quality at the Balmer-jump, 
that constrains  $\log{g}$. 
With these in mind the comparison results demonstrate a reliable spectral analysis  
of this study.

\section{Discussion and Summary}  

We selected 607 hot subdwarf candidates by 
cross-matching the catalog of Geier et al. (2019) with the  LAMOST DR6 and DR7 spectral database, 
and identified 182 hot subdwarf stars, among which 
108 stars are newly discovered. Together with 
the 682 hot subdwarf stars identified by Lei et al. (2018, 2019b), we 
found 864 hot subdwarfs in the LAMOST spectral database, and 349 of them are 
new discoveries. 

The hot subdwarf candidates in Lei et al. (2018, 2019b) were selected  
visually in the Gaia DR2 HR-diagram, which means a little different selection filter
from the one used by Geier et al. (2019, see section 3 in their study). 
Therefore, we cross-matched all the 864 hot subdwarf sars identified in our series of  
studies with the Geier et al. (2019) catalog, and found 833 common stars. 
This result demonstrate that nearly all the hot subdwarf stars identified in 
Lei et al. (2018, 2019b) are included in Geier et al. (2019) catalog.  
As described in section 2.3, 2513 candidates from the catalog of Geier et al. (2019) 
have LAMOST spectra, of which 1348 have SNR-u larger than 10, and 833 of them 
were spectroscopically identified as hot subdwarf stars. Based on these results, 
one can roughly estimate the fraction of 
hot subdwarf stars in the catalog of Geier et al. (2019).
 
Fig 5 shows the distributions of selected hot subdwarf candidates (left) 
and the fraction of confirmed hot subdwarfs (right) with respect to 
Gaia $G$ band magnitude. As showed in the left panel, for the brighter sample (e.g., 9 $<$ Gaia $G$ mag $<13$, 
that usually represents higher SNR), the candidates with SNR-u larger than 10 (blue-dashed histogram) have 
nearly the same size  as the whole sample (red-solid histogram), which 
means a good completeness of the bright end of the catalog. However,  only 
part of these stars were identified as hot subdwarf stars (green-dotted histogram). 
This result also can be seen clearly in the right panel. The fraction of candidates 
with SNR-u larger than 10 in the whole sample (grey-dashed curve) decreases from 100\% to 80\% 
within this magnitude range, and the fraction of hot subdwarf stars among  
the candidates with SNR-u larger than 10 (blue-dotted curve) is nearly the same 
as in the whole sample (red-solid curve), e.g., roughly between 10\% and 40\%. 
These results demonstrate that hot subdwarf fraction of the candidates in 
Geier et al. (2019) for brighter stars (e.g., 9 $<$ Gaia $G$ mag $<$ 13) is 
roughly from 10\% to 40\%, and increasing gradually with the magnitude. 
It can be understood that there are more 
O/B type MS stars, rather than hot subdwarf stars, in the  brighter 
part of the catalog of Geier et al. (2019) catalog. Thus, the brighter 
of the sample, much lower fraction of hot subdwarf stars it contains. 
Composite spectra were removed from our sample, if included and  
they would turn out to be real hot subdwarfs, this fraction  could be 
a little higher. 

With fainter samples (e.g.,  13 $<$ Gaia $G$ mag $<$ 16), 
the SNR-u of the spectra become lower. Therefore, many candidates do not 
enter into our sample due to low SNR, and the completeness of the 
sample becomes  worse.  
The fraction of candidates with SNR-u larger than 10 among all candidates 
drops gradually from 80\% to 60\% as the magnitude increases from 13 to 16 mag  
(see grey-dashed curve in the right panel). 
In this magnitude range, we got a hot subdwarf fraction in the 
whole sample (red-solid curve in the right panel) going from 40\% to 60\%. 
Considering that some real hot subdwarf stars were removed from our sample due to low SNR-u or composite  
feature, the hot subdwarf fraction could be higher in this magnitude range. 
At the faint end of the sample, above 16 mag, the number of candidates with SNR-u larger than 10 
drops quickly. Therefore, the candidates we analyzed in this magnitude range 
become extremely incomplete, and the hot subdwarf fraction obtained from these candidates  
is meaningless. One can expect more WDs rather than hot subdwarfs 
among the fainter candidates. 
A more accurate estimation of the fraction of hot subdwarf stars in the 
catalog of Geier et al. (2019) can be obtained when the results  
from analysis of composite spectra will be available. 

\begin{figure}
\centering
\begin{minipage}[c]{0.48\textwidth}
\includegraphics [width=88mm]{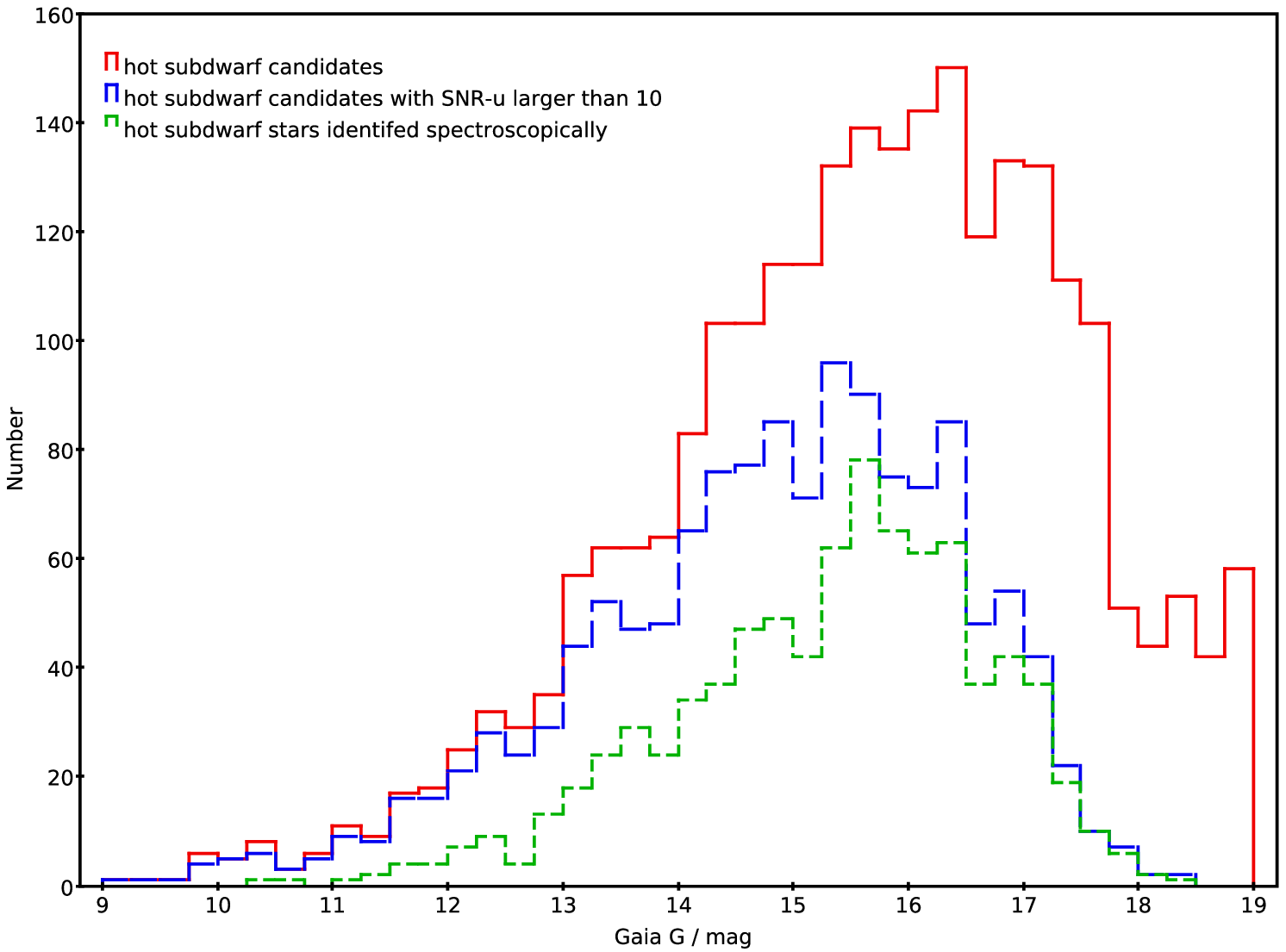}
\end{minipage}%
\begin{minipage}[c]{0.48\textwidth}
\includegraphics [width=82mm]{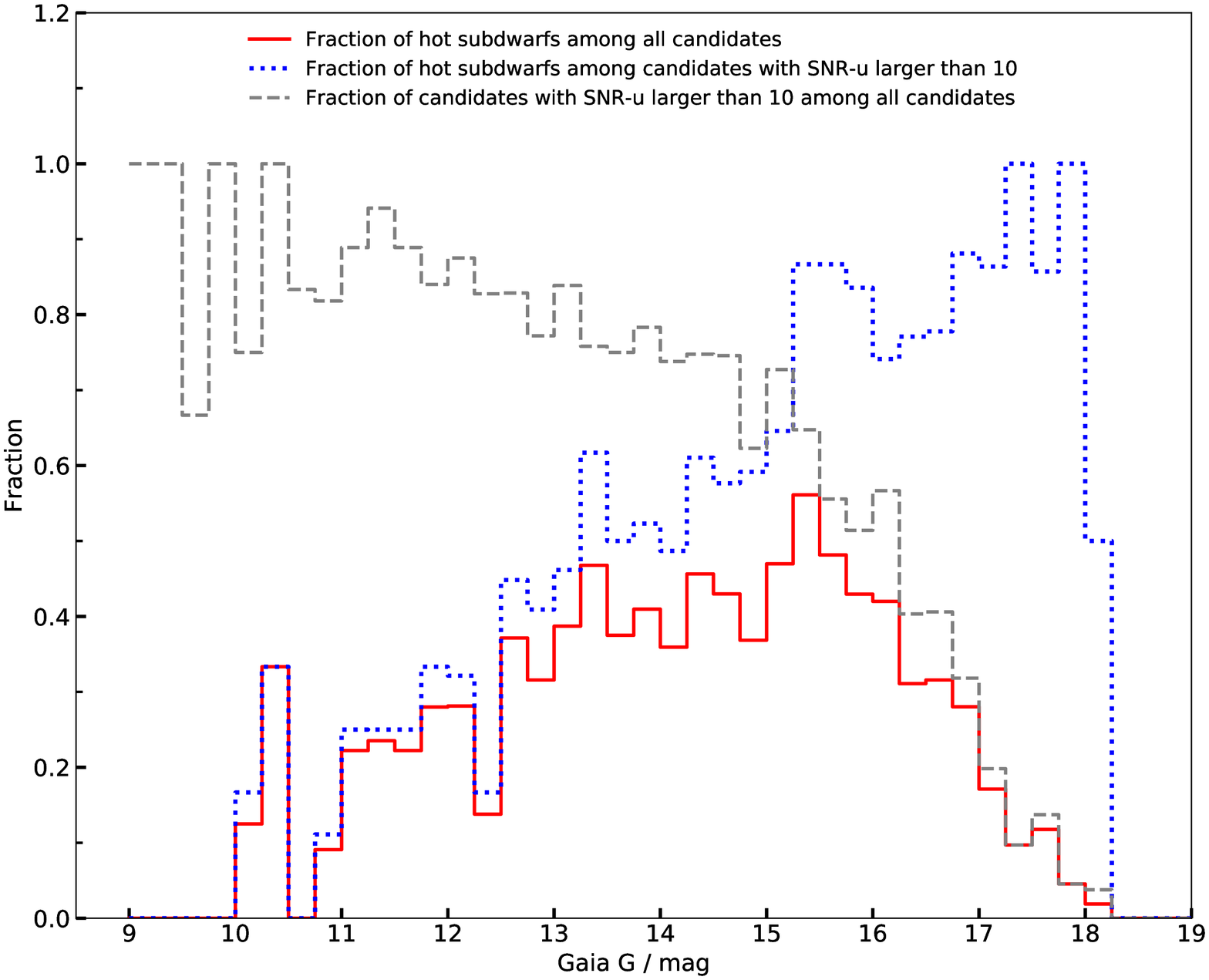} 
\end{minipage}%
\caption{ The distributions of  hot subdwarf candidates (left) 
and confirmed hot subdwarf fraction (right) with respect to Gaia $G$ magnitude. 
Left panel: red-solid histogram 
presents the distribution of 2513 candidates with a bin size of 0.25 mag, 
blue-dashed histogram denotes the distribution of 1348 candidates with SNR-u 
larger than 10, while green-dotted histogram represents the distribution of  
833 hot subdwarf stars spectroscopically identified among the 2513 candidates with LAMOST spectra. 
Right panel: red-solid curve is the hot subdwarf fraction 
among 2513 candidate stars, and blue-dotted curve is the   
hot subdwarf fraction in 1348 candidates with SNR-u larger than 10, 
while grey-dashed curve presents the fraction of candidates with 
SNR-u larger than 10 among  the 2513 candidates.}   
\end{figure} 

The results obtained in this study reflect the high efficiency of the method to  
search for hot subdwarf stars by combining  Gaia DR2 data with 
LAMOST spectra. 
We obtained reliable atmospheric parameters 
for all the hot subdwarf candidates using detailed spectral analysis with non-LTE model atmospheres. 
The atmospheric parameters are consistent with the ones from literature and  the hot subdwarf catalog of Geier et al. (2017). 
We also estimated the hot subdwarf fraction in the catalog of Geier et al. (2019) based on 
the candidates we have analyzed.
We found that the bright part (9 mag $<$ Gaia $G$ $<$ 13 mag) of the catalog 
is nearly complete, but has many false-positive candidates (over 60\%, mosttly B-type stars). 
In the  13 $<$ Gaia $G$  $<$ 16 magnitude range the hot subdwarf fraction goes from 
40\% to 60\%. The completeness of the catalog degrades quickly above $G$ = 16 mag.    
Furthermore, we selected about 150 hot subdwarf candidates with composite spectra in LAMOST 
DR6 and DR7. The results from their spectral analysis will be reported in a forthcoming paper. 
Since all spectra are observed with the same equipment and analyzed with the same method, 
we believe that the LAMOST hot subdwarf sample will make 
important contributions to study the  formation and evolution of 
these special blue objects.

\acknowledgments
We thank the anonymous referee for their valuable suggestions 
and comments which improved this work greatly. 
L.Z. acknowledges support from National Natural Science Foundation 
of China Grant No 11503016, Natural Science Foundation of Hunan province Grant No.2017JJ3283,  
the Youth Fund project of Hunan Provincial Education Department 
Grant No.15B214, Cultivation Project for LAMOST Scientific Payoff and Research Achievement of CAMS-CAS. 
This work is supported by the National Natural Science Foundation 
of China Grant Nos. 11390371, 11988101, 11973048, National 
Key R\&D Program of China No.2019YFA0405502,   
the Astronomical Big Data Joint Research
Center, co-founded by the National Astronomical Observatories, Chinese 
Academy of Sciences and the Alibaba Cloud. 
This research has used the services of \mbox{\url{www.Astroserver.org}} 
under reference D879YE and D880YE. 
P.N. acknowledges support from the Grant Agency of the Czech Republic (GA\v{C}R 18-20083S). 
The LAMOST Fellowship is supported by Special Funding for Advanced Users, 
budgeted and administered by the Center for Astronomical 
Mega-Science, Chinese Academy of Sciences (CAMS). 
Guoshoujing Telescope (the Large Sky Area Multi-Object Fiber 
Spectroscopic Telescope LAMOST) is a National Major Scientific 
Project built by the Chinese Academy of Sciences. 
Funding for the project has been provided by the 
National Development and Reform Commission. 
LAMOST is operated and managed by the National Astronomical Observatories, 
Chinese Academy of Sciences.




\begin{thebibliography}{} 
\bibitem[Amaro-Seoane et al.(2017)]{2017arXiv170200786A} Amaro-Seoane, Pau., Audley, Heather., Babak, Stanislav., et al.  \ 2017, arXiv: 1702.00786 
\bibitem[Baran et al.(2012)]{2012MNRAS.424.2686B} Baran, A. S., Reed, M. D., Stello, D.,  et al.\ 2012, \mnras, 424, 2686  
\bibitem[Battich et al.(2018)]{2018A&A...614A.136B} Battich, Tiara., Bertolami, Marcelo M. Miller., C\'orsico, Alejandro H.,  et al.\ 2018, \aap, 614, 136  
\bibitem[Bu et al.(2017)]{2017ApJS..233....2B} Bu, Yude., Lei, Zhenxin., Zhao, Gang., et al. \ 2017, \apjs, 233, 2 
\bibitem[Bu et al.(2019)]{2019ApJ...886..128B } Bu, Yude., Zeng, Jingjing., Lei, Zhenxin., et al. \ 2019, \apj, 886, 128 
\bibitem[Byrne et al.(2018)]{2018MNRAS.475.4728B} Byrne, Conor M., Jeffery, C. Simon., Tout, Christopher A., et al.\ 2018, \mnras, 475, 4728 
\bibitem[Charpinet et al.(2011)]{ 2011A&A...530A...3C} Charpinet, S., Van Grootel, V., Fontaine, G.,  et al.\ 2011, \aap, 530, 3 
\bibitem[Chen et al.(2013)]{2013MNRAS.434..186C} Chen, Xuefei., Han, Zhanwen., Deca, Jan.,  et al.\ 2013, \mnras, 434, 186
\bibitem[Copperwheat et al.(2011)]{2011MNRAS.415.1381C} Copperwheat, C. M., Morales-Rueda, L., Marsh, T. R.,  et al.\ 2011, \mnras, 415, 1381
\bibitem[Cui et al.(2012)]{2012RAA....12.1197C} Cui, Xiang-Qun., Zhao, Yong-Heng., Chu, Yao-Quan., et al.\ 2012, RAA, 12, 1197
\bibitem[Dorman et al.(1993)]{1993ApJ...419..596D} Dorman, Ben., Rood, Robert T., \& O'Connell, Robert W.\ 1993, \apj, 419, 596
\bibitem[Edelmann et al.(2003)]{2003A&A...400..939E} Edelmann, H., Heber, U., Hagen, H.-J.,  et al.\ 2003, \aap, 400, 939
\bibitem[Gaia Collaboration et al.(2018)]{2018A&A...616A...1G} Gaia Collaboration, Brown, A., Vallenari, A., et al. 2018,  \aap, 616, 1 
\bibitem[Geier et al.(2013)]{2013A&A...557A.122G} Geier, S., Heber, U., Edelmann, H., et al.\ 2013, \aap, 557, 122
\bibitem[Geier et al.(2015)]{2015Sci...347.1126G} Geier, S., Fürst, F., Ziegerer, E., et al.\ 2015, Science, 347, 1126
\bibitem[Geier et al.(2017)]{2017A&A...600A..50G} Geier, S., $\O$stensen, R. H., Nemeth, P., et al.\ 2017, \aap, 600, 50
\bibitem[Geier et al.(2019)]{2019A&A...621A..38G} Geier, S.,  Raddi, R., Gentile Fusillo,  N. P., et al.\ 2019, \aap, 621, 38
\bibitem[Hagen et al.(1995)]{1995A&AS..111..195H} Hagen, H. -J., Groote, D., Engels, D., et al.\ 1995, \aaps, 111, 195
\bibitem[Han et al.(2002)]{2002MNRAS.336..449H} Han, Z., Podsiadlowski, Ph., Maxted, P. F. L., et al.\ 2002, \mnras, 336, 449
\bibitem[Han et al.(2003)]{2003MNRAS.341..669H} Han, Z., Podsiadlowski, Ph., Maxted, P. F. L., et al. \ 2003, \mnras, 341, 669
\bibitem[Heber(1986)]{1986A&A...155...33H } Heber, U.\ 1986, \aap, 155, 33    
\bibitem[Heber(2009)]{2009ARA&A..47..211H} Heber, U.\ 2009, \araa, 47, 211 
\bibitem[Heber(2016)]{2016PASP..128h2001H} Heber, U.\ 2016, \pasp, 128, 2001 
\bibitem[Hubeny \& Lanz(2017)]{2017arXiv170601859H} Hubeny, I., \& Lanz, T.\ 2017, arXiv:1706.01859
\bibitem[Jeffery et al.(2017)]{2017MNRAS.465.3101J} Jeffery, C.~S., Baran, A.~S., Behara, N.~T., et al.\ 2017, \mnras, 465, 3101 
\bibitem[Kawaler et al.(2010)]{ 2010MNRAS.409.1509K} Kawaler, S. D., Reed, M. D., $\O$stensen, R. H., et al.\ 2010, \mnras, 409, 1509  
\bibitem[Kawka et al.(2015)]{2015MNRAS.450.3514K} Kawka, A., Vennes, S., O'Toole, S., et al.\ 2015, \mnras, 450, 3514 
\bibitem[Kepler et al.(2015)]{2015MNRAS.446.4078K} Kepler, S. O., Pelisoli, I., Koester, D., et al.\ 2015, \mnras, 446, 4078
\bibitem[Kepler et al.(2016)]{2016MNRAS.455.3413K} Kepler, S. O., Pelisoli, I., Koester, D.,  et al.\ 2016, \mnras, 455, 3413
\bibitem[Kilkenny et al. (1988)]{1988SAAOC..12....1K} Kilkenny, D., Heber, U., \& Drilling, J. S. \ 1988, SAAOC, 12, 1 
\bibitem[Kilkenny et al. (1997)]{1997MNRAS.287..867K} Kilkenny, D., O'Donoghue, D., Koen, C., et al. \ 1997, \mnras, 287, 867 
\bibitem[Kupfer et al. (2015)]{2015A&A...576A..44K} Kupfer, T., Geier, S., Heber, U., et al. \ 2015, \aap, 576, 44
\bibitem[Kupfer et al. (2018)]{2018MNRAS.480..302K} Kupfer, T., Korol, V., Shah, S., et al. \ 2018, \mnras, 480, 302 
\bibitem[Lanz \& Hubeny.(2007)]{2007ApJS..169...83L} Lanz, Thierry., \&  Hubeny, Ivan. \ 2007, \apjs, 169, 83 
\bibitem[Latour et al. (2014]{2014ApJ...795..106L} Latour, M., Randall, S. K., Fontaine, G.,  et al.\ 2014, \apj, 795, 106 
\bibitem[Latour et al.(2018]{2018A&A...618A..15L}Latour, Marilyn., Randall, Suzanna K., Calamida, Annalisa.,  et al.\ 2018, \aap, 618, 15 
\bibitem[Lei et al.(2015)]{2015MNRAS.449.2741L} Lei, Zhenxin., Chen, Xuemei., Zhang, Fenghui., et al.\ 2015, \mnras, 449, 2741
\bibitem[Lei et al.(2016)]{2016MNRAS.463.3449L} Lei, Zhenxin., Zhao, Gang., Zeng, Aihua.,  et al.\ 2016, \mnras, 463, 3449 
\bibitem[Lei et al.(2018)]{2018ApJ...868...70L} Lei, Zhenxin., Zhao, Jingkun., N\'emeth, P\'eter.,  et al.\ 2018, \apj, 868, 70
\bibitem[Lei et al.(2019a)]{2019PASJ...71...41L} Lei, Zhenxin., Bu, Yude., Zhao, Jingkun., et al.\ 2019a, \pasj, 71, 41   
\bibitem[Lei et al.(2019b)]{2019ApJ...881..135L} Lei, Zhenxin., Zhao, Jingkun., N\'emeth, P\'eter.,  et al.\ 2019b, \apj, 881, 135 
\bibitem[Li et al.(2018)]{2018AJ....156...87L} Li, Yin-Bi., Luo, A-Li., Zhao, Gang., et al.\ 2018, \aj, 156, 87 
\bibitem[Luo et al.(2016a)]{2016CQGra..33c5010L } Luo, Jun., Chen, Li-Sheng., Duan, Hui-Zong.,  et al.\ 2016a, CQGra, 33, 5010
\bibitem[Luo et al.(2016b)]{2016ApJ...818..202L} Luo, Yang-Ping., N\'emeth, P., Liu, Chao.,  et al.\ 2016b, \apj, 818, 202
\bibitem[Luo et al.(2019)]{2019ApJ...881....7L} Luo, Yang-Ping., N\'emeth, P., Deng, Licai.,  et al.\ 2019, \apj, 881, 7 
\bibitem[Maxted et al.(2001)]{2001MNRAS.326.1391M} Maxted, P. F. L., Heber, U., Marsh, T. R., et al.\ 2001, \mnras, 326, 1391 
\bibitem[Mickaelian (2008)]{2008AJ....136..946M} Mickaelian, A. M.\ 2008, \aj, 136, 946  
\bibitem[Mickaelian et al.(2007)]{2007A&A...464.1177M} Mickaelian, A. M., Nesci, R., Rossi, C., et al.\ 2007, \aap, 464, 1177 
\bibitem[Moehler et al.(1990)]{1990A&AS...86...53M} Moehler, S., Richtler, T., de Boer, K. S.,  et al.\ 1990, \aaps, 86, 53
\bibitem[Moehler et al.(2014)]{2014A&A...565A.100M} Moehler, S., Dreizler, S., LeBlanc, F.,  et al.\ 2014, \aap, 565, 100
\bibitem[N\'emeth et al.(2014)]{ 2014ASPC..481...95N} N\'emeth, P., Östensen, R., Tremblay, P., et al.\ 2014, ASPC, 481, 95
\bibitem[N\'emeth et al. (2012)]{2012MNRAS.427.2180N} N\'emeth, P., Kawka, A., \&  Vennes, S.\ 2012, \mnras, 427, 2180
\bibitem[N\'emeth (2017)]{2017OAst...26..280N} N\'emeth, P. \ 2017, Open Astronomy, 26, 280  
\bibitem[Napiwotzki et al.(2004)]{2004Ap&SS.291..321N} Napiwotzki, R., Karl, C. A., Lisker, T.,  et al.\ 2001, \apss, 291, 321
\bibitem[Naslim et al.(2013)]{2013MNRAS.434.1920N} Naslim, N., Jeffery, C.S., Hibbert, A., et al.\ 2013, \mnras, 434, 1920
\bibitem[Naslim et al.(2019)]{2019arXiv191008947N} Naslim, N., Jeffery, C.S., \& Woolf, V. M. \ 2019, arXiv:1910.08947 
\bibitem[Paczy\'nski(1971)]{1971AcA....21....1P} Paczy\'nski, B.\ 1971, Acta Astron, 21, 1 
\bibitem[Raddi et al.(2019)]{2019MNRAS.489.1489R} Raddi, R., Hollands, M. A., Koester, D., et al. \ 2019, \mnras, 489, 1489 
\bibitem[Vennes, Kawka \& Németh(2011)]{2011MNRAS.410.2095V} Vennes, S., Kawka, A \& N\'emeth, P. \ 2011, \mnras, 410, 2095
\bibitem[Vennes et al.(2017)]{2017Sci...357..680V} Vennes, S., Nemeth, P., Kawka, A., et al. \ 2017, Science, 357, 680 
\bibitem[Vos et al.(2019)]{2019MNRAS.482.4592V} Vos, Joris., Vu$\check\mathrm{c}$kovi\'c, Maja., Chen, Xuefei., et al.  \ 2019, \mnras, 482, 4592
\bibitem[Wang et al.(2009)]{2009MNRAS.395..847W} Wang, B., Meng, X., Chen, X., et al.\ 2009, \mnras, 395, 847
\bibitem[Wisotzki et al.(1996)]{1996A&AS..115..227W} Wisotzki, L., Koehler, T., Groote, D., et al.\ 1996, \aaps, 115, 227 
\bibitem[Zhang \& Jeffery(2012)]{2012MNRAS.419..452Z} Zhang, Xianfei., \&  Jeffery, C. S.\ 2012, \mnras, 419, 452
\bibitem[Zhang et al.(2017)]{2017ApJ...835..242Z} Zhang, Xianfei., Hall, Philip D., Jeffery, C. Simon., et al.\ 2017, \apj, 835, 242
\bibitem[Zhao et al.(2006)]{2006ChJAA...6..265Z} 	Zhao, Gang., Chen, Yu-Qin., Shi, Jian-Rong., et al.\ 2006, ChJAA, 6, 265  
\bibitem[Zhao et al.(2012)]{2012RAA....12..723Z} Zhao, Gang., Zhao, Yong-Heng., Chu, Yao-Quan., et al.\ 2012, RAA, 12, 723 
\bibitem[Zong et al.(2018]{2018ApJ...853...98Z} Zong, Weikai.,  Charpinet, St\'ephane.,  Fu, Jian-Ning., et al.\ 2018, \apj, 835, 98 

\end{thebibliography}
\end{document}